%% file: main.tex
\documentclass[sigconf, nonacm]{acmart}

\setcopyright{none}
\acmConference[ISCA '27]{54th IEEE/ACM International Symposium on Computer Architecture}{June 5--9, 2027}{Atlanta, Georgia, USA}
\acmYear{2027}
\acmISBN{978-x-xxxx-xxxx-x/YY/MM}
\acmDOI{XXXXXXX.XXXXXXX}
\usepackage{amsmath,amsfonts}
\usepackage{booktabs}
\usepackage{algorithmic}
\usepackage{graphicx}
\usepackage{textcomp}
\usepackage{xcolor}
\usepackage{algorithmic}
\usepackage{textcomp}
\usepackage{xcolor}
\usepackage{fancyhdr}
\usepackage{microtype}[final]
\usepackage[normalem]{ulem}
\usepackage{tikz}
\usepackage{tcolorbox}
\usepackage[linesnumbered,ruled,vlined]{algorithm2e}
\usepackage{multirow}
\usepackage{colortbl}
\usepackage[bottom]{footmisc}
\usepackage{stfloats}
\usepackage{soulutf8}
\usepackage{enumitem}
\usepackage{multicol}
\usepackage{qcircuit}
\usepackage{fvextra}
\usepackage{forest}
\usepackage{booktabs}
\usepackage{listings}
\usepackage[labelformat=simple]{subcaption}

\renewcommand{\theenumii}{\theenumi.\arabic{enumii}.}

\newcommand{\subheading}[1]{\noindent \textbf{#1:}}

\newcommand{\ours}[0]{\texttt{Augur}}

\newcommand{\rev}[1]{\textcolor{black}{#1}}

\definecolor{aliceblue}{rgb}{0.97, 0.96, 1.0}

\newtcolorbox{boxK}{
    sharpish corners, 
    boxrule = 0pt,
    toprule = 1pt, 
    enhanced,
    fuzzy shadow = {0pt}{-2pt}{-0.5pt}{0.5pt}{black!35} 
}

\newtcolorbox{boxH}{
    sharpish corners,
    boxrule = 0pt, 
    leftrule = 2pt 
}

\makeatletter
\let\MYcaption\@makecaption
\makeatother

\usepackage[font=footnotesize,labelformat=simple]{subcaption}

\makeatletter
\let\@makecaption\MYcaption
\makeatother

\keywords{Data Prefetching, Next-Generation Reservoir Computing.}

\begin{document}

\title{Augur: An Online Learning-based Hardware Data Prefetcher using Reservoir Computing}
\author[1]{Satvik Maurya$^\dagger$, David Trilla$^\ddagger$, Swamit Tannu$^\dagger$, Alper Buyuktosunoglu$^\ddagger$}
\author[2]{University of Wisconsin-Madison$^\dagger$, IBM Research$^\ddagger$}


\begin{abstract}

Machine-learning-based hardware prefetchers can learn complex access patterns, but practical deployment demands online adaptation, low-latency inference, and modest hardware cost. We present Augur, a hardware delta prefetcher based on next-generation reservoir computing (NGRC) that meets all three requirements.
Augur formulates delta prediction as a history-based regression problem. It constructs a fixed-size reservoir from hashed delta features over a short context window and predicts the next delta using a single dot product. The model is updated online via one-step SGD and requires no offline pre-training, no per-benchmark tuning, and no floating-point arithmetic. Replacing the integer multiplications traditionally used to generate nonlinear reservoir features with bitwise AND operations removes all feature-generation multipliers with no loss in performance. Because Augur regresses a continuous delta, it needs no action set chosen offline for a fixed set of workloads. Augur pairs the reservoir with a simple Markov-based delta predictor for simple delta patterns, and with a selective issuance policy that filters and throttles wasteful prefetches; this policy accounts for about 40\% of Augur's advantage and also improves Pythia. Evaluated on 169 traces spanning GAP, Ligra, PARSEC, SPEC CPU 2006, SPEC CPU 2017, and AI workloads, Augur outperforms the state-of-the-art ML-based delta prefetcher by 2.9\%, 4.5\%, and 5.3\% in geomean speedup on 1-, 4-, and 8-core systems, respectively, while achieving $\sim$2× higher prefetch accuracy and 37.8 percentage points fewer overpredictions. Augur is practical to implement at 33 KB storage, 0.32 mm² area, and 32.7 mW in 12 nm.

\end{abstract}

\maketitle

\input{inputs/1Introduction}

\input{inputs/2Motivation}
\input{inputs/3Background}
\input{inputs/4Design}
\input{inputs/5Design2}
\input{inputs/6Methodology}
\input{inputs/7Results}
\input{inputs/8Discussion}
\input{inputs/9RelatedWorks} 
\input{inputs/10Conclusions}

\bibliographystyle{ACM-Reference-Format}
\bibliography{refs}

\end{document}

%% file: inputs/1Introduction.tex
\section{Introduction}


The rapidly widening gap between processing and memory speeds, the ``Memory Wall''~\cite{wulf1995hitting}, remains one of the most pressing bottlenecks in modern computing.
Hardware data prefetching, which speculatively fetches data into the caches before it is demanded, is a first-order technique for mitigating this gap.
A data prefetcher answers the question: what address will the program access next?
Strategies fall broadly into two families: temporal and delta/stride prefetchers.
Temporal prefetchers \emph{memorize} and replay address sequences, exploiting temporal locality.
Delta prefetchers predict the \emph{difference} (delta) between consecutive cache-line addresses accessed by the same PC; because deltas are compact integers rather than full addresses, delta prefetchers are inherently more storage-efficient and practical than temporal prefetchers.
We focus on delta prefetching.

Traditional delta prefetchers rely on heuristics: stride detectors~\cite{stride} detect PCs with constant deltas, the spatial pattern matcher SMS~\cite{sms} exploits spatial locality within memory regions, Markov predictors~\cite{markov} capture short delta patterns, the signature-based SPP~\cite{spp} uses hashed delta histories, and Berti~\cite{navarro2022berti} focuses on timeliness; others include VLDP~\cite{vldp}, Bingo~\cite{bingo}, IPCP~\cite{ipcp}, and DSPatch~\cite{dspatch}.
They work well for regular access patterns but struggle with the complex, program-dependent locality of many modern workloads.

Machine learning (ML) offers a compelling alternative, discovering patterns and correlations hidden to hand-crafted heuristics.
On the temporal side, Voyager~\cite{Shi2021} uses an LSTM with attention-based embeddings to detect repeating address sequences, while Twilight~\cite{duong2024new} clusters addresses offline and reformulates temporal prefetching as classification.
On the delta side, proposals frame delta prediction as a natural language processing problem using RNNs, LSTMs, or transformers~\cite{hashemi2018learning, zeng2017long, Srivastava2019, shineural}; as a reinforcement learning (RL) problem~\cite{pythia, peled_rl, zhang2022resemble}; or via customized neural network architectures including spiking neural networks~\cite{pathfinder, peled2018neural}.

Current state-of-the-art ML-based delta prefetchers have one or more of the following limitations: 
\begin{enumerate}[leftmargin=*, topsep=4pt, itemsep=2pt]
    \item \textbf{Offline training and fragile generalization:} Nearly every ML-based prefetcher trains its model offline on program traces before deployment, restricting adaptation to unseen workloads.
    \item \textbf{Expensive inference:} Approaches built on LSTMs, transformers, or deep neural networks demand thousands of integer or floating-point operations per prediction. 
    \item \textbf{\rev{Offline-chosen action sets}:} RL-based prefetchers such as Pythia rely on a limited, pre-defined set of deltas chosen through offline workload simulations, \rev{which ties the prefetcher to the workloads used to choose that set}. 
\end{enumerate}


\begin{center}
    In this paper, we ask-- \emph{can a hardware delta prefetcher retain the pattern-learning power of ML while training entirely online, support cheap inference, and require no workload-specific offline tuning?}
\end{center}

We answer affirmatively with \textbf{\ours}, a level-2 cache (L2C) delta prefetcher based on \emph{next-generation reservoir computing} (NGRC)~\cite{Gauthier2021}.
NGRC maintains a compact context window of input features---in our case, the PC-hashed delta and a rolling delta signature---and generates non-linear features by forming \emph{monomials} (products of features) from this window.
The delta signature is the PC XORed with a rolling XOR-shift hash of the per-PC delta history, encoding the temporal sequence of recent deltas observed for a given PC.
The concatenation of linear and non-linear features constitutes a \emph{reservoir}, and predicting the next delta reduces to a single dot product between the reservoir and a learned weight vector.
\rev{Because \ours{} formulates delta prediction as regression rather than classification, it predicts
deltas in $[-64, 63]$ without an action set chosen offline (Section~\ref{sec:sources}).}
Training is a one-step stochastic gradient descent (SGD) update per demand access, so \ours~is fully online with no offline pre-training, \rev{no per-benchmark hyperparameter tuning}, and no floating-point arithmetic.
\rev{To keep the nonlinear features cheap}, we replace the integer multiplications traditionally used to generate them with \emph{bitwise AND} operations.
\rev{This removes the 528 multipliers of nonlinear feature generation and yields a
fixed-width result in a single cycle with no overflow, at no loss in performance
(Section~\ref{sec:sources}).}
\rev{The nonlinear features help on PCs whose next delta depends on the combined context, but they are optional: a linear-only \ours~is only 0.3\% slower.}
\ours~is a \emph{tournament prefetcher}: a lightweight Markov table handles trivial, repetitive delta patterns (e.g., constant or alternating deltas) at near-zero cost, \rev{the reservoir handles the patterns the Markov table cannot, and a throttled next-line prefetcher covers PCs that neither predicts confidently.}

\vspace{0.05in}

\subheading{\rev{Selective issuance} through filtering and throttling}
\ours's central design philosophy is that \emph{not prefetching} is often better than prefetching incorrectly.
Useless prefetches evict useful cache lines, consume scarce memory bandwidth, and contend for MSHR and prefetch queue entries---problems that compound in multi-core systems with shared bandwidth.
\ours~therefore pairs its predictors with a \rev{selective issuance policy}.
First, per-PC usefulness counters\rev{, kept separately for predicted and next-line prefetches,} track the ratio of useful to issued prefetches; when a PC's ratio drops, its \rev{prefetch degree is reduced} or its prefetching is disabled entirely.
Second, a direct-mapped prefetch filter keeps recently prefetched or cache-resident lines from flooding the prefetch queue.
\rev{Usefulness feedback and prefetch filtering have been used before~\cite{fdp, spp, ppf}; what is new in \ours~is tracking usefulness per PC and per source, and that the policy does not depend on the predictor: applied to Pythia, it improves Pythia by 6.4\%.}
\rev{\ours~also caps the degree when the prefetch queue or DRAM bandwidth nears saturation, and uses an adaptive lookahead based on the prefetch fill latency and each PC's inter-access gap.}

 \begin{figure}[t]
    \centering
    \includegraphics[width=\linewidth]{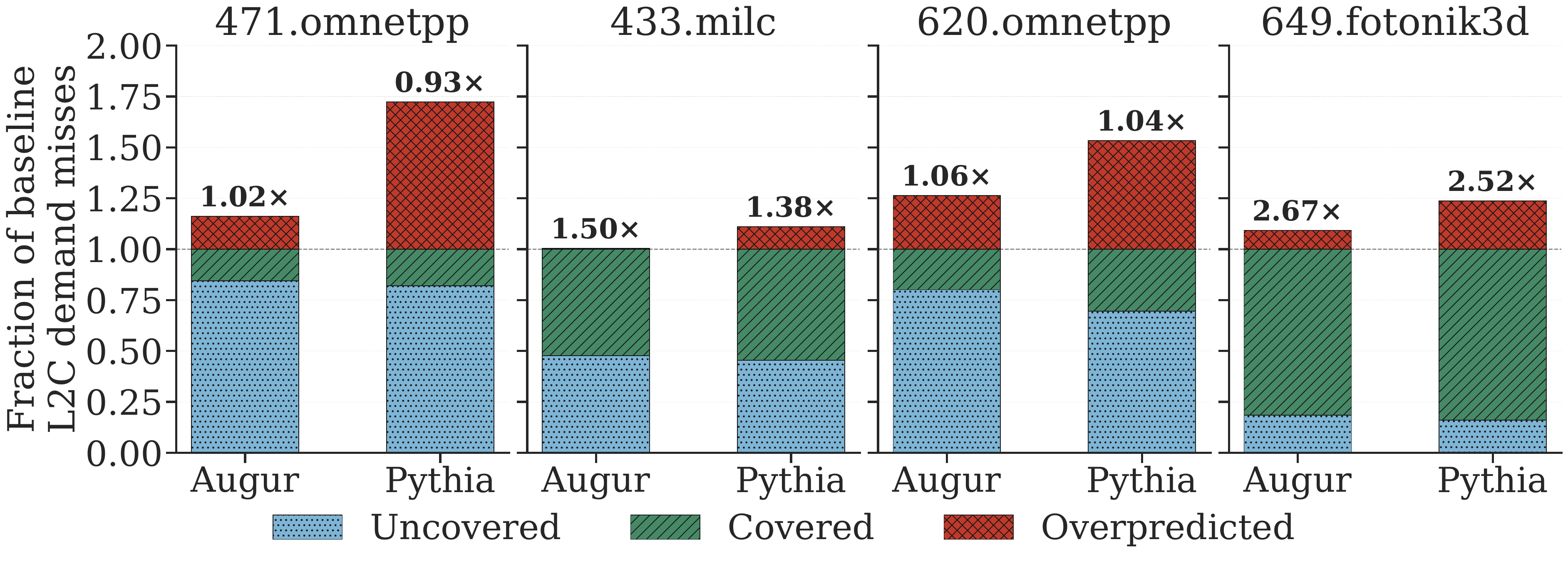}
    \caption{Coverage and overpredictions by \ours~and Pythia for representative SPEC 2006/2017 workloads. Speedups over a no-prefetch baseline are annotated above the bars.}
    \label{fig:intro_cov}
    \vspace{-0.15in}
\end{figure}

The result is a selective rather than aggressively high-coverage prefetcher: for four SPEC 2006 and SPEC 2017 workloads (Figure~\ref{fig:intro_cov}), \ours's coverage is \emph{lower} than Pythia's.
However, Pythia's coverage costs more overpredictions, which pollute caches and waste DRAM bandwidth, yielding lower speedups than \ours.
\ours~converts fewer, high-quality prefetches into greater speedup\rev{, and its advantage grows from one to eight cores}.

This paper makes three primary contributions:

\begin{enumerate}[leftmargin=*, topsep=2pt, itemsep=0pt]
    \item We introduce NGRC as an ML framework for hardware data prefetching that naturally satisfies the constraints of online learning, \rev{cheap} inference, and bounded storage that have limited the practicality of prior ML-based prefetchers. \rev{We make its nonlinear features multiplier-free by replacing integer multiplications with bitwise AND operations, at no loss in performance, and show when these features can be omitted altogether.}

    \item \rev{We propose a selective issuance policy that tracks prefetch usefulness per PC and per source and filters redundant prefetches. The policy is independent of the predictor: it accounts for about 40\% of \ours's advantage over Pythia, and improves Pythia itself by 6.4\%.}

    \item We design \textbf{\ours}, an L2C delta prefetcher that pairs NGRC with a Markov predictor \rev{and a throttled next-line prefetcher} in a tournament architecture\rev{, and show how much of its speedup comes from prediction and how much from selective issuance}. 
\end{enumerate}

\subheading{Results} We evaluate \ours~on 169 traces spanning GAP~\cite{gap}, Ligra~\cite{ligra}, PARSEC~\cite{parsec}, SPEC CPU 2006~\cite{spec2006}, SPEC CPU 2017~\cite{spec2017}, and AI workloads. \ours~outperforms the state-of-the-art ML-based prefetcher Pythia by \textbf{2.9\%, 4.5\%, and 5.3\%} in geomean speedup across 1-, 4-, and 8-core configurations, with $\sim$\textbf{2$\times$ higher prefetch accuracy} and \rev{37.8 percentage points} fewer overpredictions---all without any suite-specific offline tuning. \ours~is robust across DRAM bandwidth and LLC capacity configurations, and a full RTL implementation synthesized in a commercial 12\,nm process confirms its practicality at \rev{33\,KB} storage, 0.32\,mm\textsuperscript{2} area, and 32.7\,mW power.

%% file: inputs/2Motivation.tex
\section{Machine Learning for Data Prefetching}
\label{sec:motivation}

We highlight the limitations of current approaches and show the need for performant yet practical ML-based data prefetchers.

\subsection{Current state-of-the-art approaches}

ML models' ability to learn complex patterns and correlations makes them a natural fit for data prefetching.
ML-based prefetchers fall broadly into (i) temporal and (ii) spatial (delta) prefetchers.

\subsubsection{Temporal prefetchers}
Temporal prefetchers learn repeating address sequences, exploiting temporal locality.
For example, if a PC sees memory addresses $A \rightarrow B \rightarrow F \rightarrow D$, the prefetcher can memorize this sequence and replay it the next time that PC encounters address $A$.
State-of-the-art temporal prefetchers include Voyager~\cite{Shi2021} and Twilight~\cite{duong2024new}.
Voyager uses a long short-term memory (LSTM) model with attention-based embeddings to detect temporal locality and issue accurate, high-coverage prefetches.
Twilight refines Voyager by observing that an address is only ever followed by a finite set of addresses, which lets it \emph{cluster} addresses (trained offline) and formulate temporal prefetching as a classification problem.

\subsubsection{Delta prefetchers}
Delta prefetchers predict the delta between the current and next address, exploiting the recurrence of deltas due to \emph{spatial} locality.
ML-based delta prefetchers~\cite{hashemi2018learning,zhang2022resemble,Srivastava2019,zeng2017long,peled_rl,peled2018neural,pathfinder,pythia,Srivastava2020,Zhang2020,shineural} formulate delta prefetching as (a) a natural language processing problem using recurrent neural networks (RNNs), LSTMs or transformers~\cite{hashemi2018learning,Zhang2020,zeng2017long,peled2018neural,Srivastava2019,Srivastava2020,shineural}; (b) a reinforcement learning problem~\cite{pythia,yang2025reinforcement,zhang2022resemble,peled_rl}; or (c) customized standard or spiking neural networks~\cite{peled2018neural,pathfinder}.

\subsection{Limitations of current approaches}
These ML-based prefetchers improve on their non-ML counterparts: Pythia~\cite{pythia} outperforms non-ML state-of-the-art prefetchers by up to 20.3\%, while Twilight~\cite{duong2024new} outperforms delta and temporal prefetchers like IPCP~\cite{ipcp}, Domino~\cite{domino}, Pythia~\cite{pythia}, Voyager~\cite{Shi2021}, and STMS~\cite{somogyi_stems} by up to 5.8\%. 
However, all of them have impracticalities, discussed below.

\subsubsection{Training and limited robustness}
Almost every ML-based prefetcher is trained offline on program traces before it can make real-time predictions, which often prevents its performance from generalizing to unseen programs; training LSTMs or transformers is also computationally intensive by itself.
RL-based prefetchers like Pythia are an exception, but Pythia required hundreds of offline simulations to find the best-performing feature set for its RL agent, which may not generalize beyond the benchmarks used. \rev{Pythia's hardware uses fixed-point adders and multipliers for its Q-value updates~\cite{pythia}; it does
less arithmetic per access than \ours{} but searches a 256-entry evaluation queue on every demand and
holds each prediction until its reward arrives.}

\subsubsection{Inference complexity}
Most ML-based prefetchers also incur non-trivial hardware overheads for inference.
For example, Twilight~\cite{duong2024new} in its most compact, hardware-friendly form (T-LITE) requires 57K integer operations (IOPs) for the matrix multiplications of inference; approaches using large fully connected layers or LSTMs are just as complex and compute-intensive, making them impractical.
Pythia remains the most scalable and practical ML-based prefetcher due to its lightweight table-based implementation and simple RL-based training/inference.  

\subsection{Our goal and proposal}
Our goal is an ML-based prefetcher that exploits the learning capabilities of ML without these limitations.
An ideal prefetcher should (1) learn patterns online while instructions are being executed, (2) require no workload-specific offline tuning, (3) make timely and accurate prefetches, (4) attain a high coverage, and (5) be implemented efficiently in hardware.
The following sections introduce reservoir computing and show how our reservoir-computing-based prefetcher, \ours, can outperform state-of-the-art ML-based prefetchers like Pythia across multiple benchmark suites.

%% file: inputs/3Background.tex
\section{Augur: Design}
\label{sec:augur}
\label{sec:reservoir}
In this section, we briefly describe next-generation reservoir computing (NGRC) and \rev{how} it is used by \ours{}.




\subsection{Next-generation reservoir computing}
\begin{figure}[t]
    \centering
    \includegraphics[width=\linewidth]{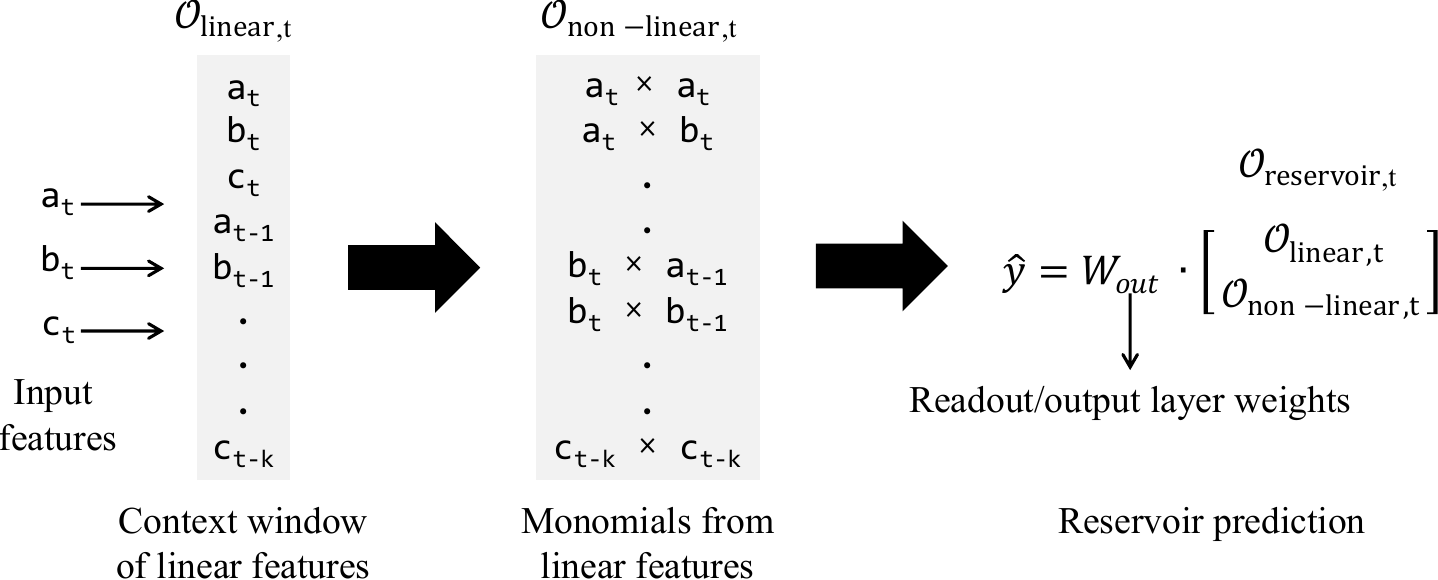}
    \caption{
    Next-generation reservoir computing. 
    }
    \label{fig:reservoir_bg}
    \vspace{-0.15in}
\end{figure}

Figure~\ref{fig:reservoir_bg} shows an NGRC predictor. $a_\mathrm{t}, b_\mathrm{t}, c_\mathrm{t}$ represent input features at time-step $t$.
A $k$-entry time-delayed history of $f(=3)$ linear features $\mathcal{O}_{\mathrm{linear}} = \{a_\mathrm{t}, b_\mathrm{t}, ..., c_{\mathrm{t-k}}\}$ generates non-linear features $\mathcal{O}_{\mathrm{non-linear}}$ by multiplying combinations of $d$ linear features. With $N=k\times f$ linear features, degree-$d$ generation yields $\binom{N+d-1}{d}$ non-linear features (Figure~\ref{fig:reservoir_bg} shows $d=2$).
Finally, the reservoir state $\mathcal{O}_{\mathrm{reservoir}}$ (a concatenation of $\mathcal{O}_{\mathrm{linear}}, \mathcal{O}_{\mathrm{non-linear}}$) yields a prediction $\hat{y}$ as the dot-product of $W_{\mathrm{out}}$ and $\mathcal{O}_{\mathrm{reservoir}}$.
During training, $W_{\mathrm{out}}$ can be determined using ridge regression~\cite{vogel2002computational,Gauthier2021} or stochastic gradient descent (SGD) using least means squares (LMS)~\cite{NIPS2002_426f990b}.

NGRC works well for time-series prediction because it makes the problem's temporal structure explicit instead of asking a large recurrent network to discover it implicitly. It turns prediction into a compact nonlinear autoregression: a short window of recent history provides memory of how the signal has evolved over the last few steps, low-order nonlinear combinations of this history provide expressiveness by capturing interactions among recent events rather than only linear trends, and a final linear readout learns which of these patterns matter for forecasting the next value. This fits data prefetching naturally, since the next delta is often determined by short-range temporal context and interactions among a few recent accesses, which NGRC can model accurately without a complex ML model.

The key advantages of NGRC for microarchitectural deployment are its support for online training via simple incremental updates, and its inference path, which reduces to a single dot product over a fixed feature vector. This contrasts sharply with other ML approaches using multi-layer perceptrons (MLPs), which require multi-layer matrix multiplications and non-linear activations at inference time, and gradient backpropagation for training — both prohibitive in area- and power-constrained settings. In the subsequent sections, we present \ours, an NGRC-based L2 delta prefetcher that predicts the next cache-line delta for a load PC and issues the corresponding prefetches, assuming 64B cache lines.

%% file: inputs/4Design.tex

\begin{figure}[t]
    \centering
    \includegraphics[width=\linewidth]{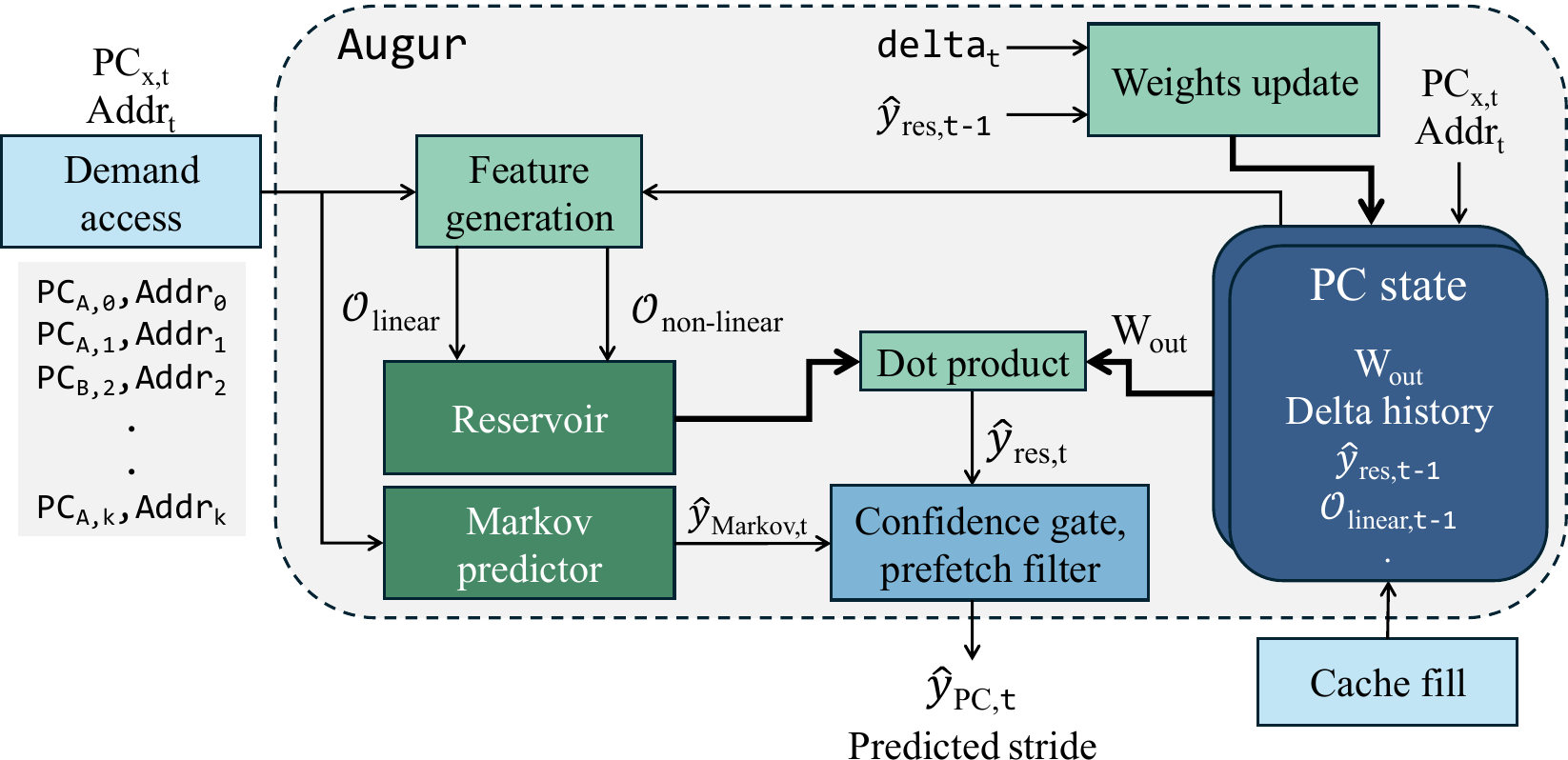}
    \caption{Overview of \ours.
    }
    \label{fig:overview}
    \vspace{-0.15in}
\end{figure}

\subsection{End-to-end architecture of Augur}
\label{sec:overview}
Figure~\ref{fig:overview} shows an overview of \ours.
On every demand access at time-step $t$, the corresponding $\mathrm{PC}_\mathrm{x,t}$ and
demand address $\mathrm{Addr}_\mathrm{t}$ are passed into \ours.
The observed delta at time-step $\mathtt{t}$ $\mathtt{delta}_\mathtt{t}$ is
computed and used to generate linear features $\mathcal{O}_\mathrm{linear}$ and non-linear
features $\mathcal{O}_\mathrm{non-linear}$, which together constitute the reservoir state
$\mathcal{O}_\mathrm{reservoir}$.
A dot product of $\mathcal{O}_\mathrm{reservoir}$ with a per-PC weight vector $W_\mathrm{out}$
yields the reservoir's predicted next delta $\hat{y}_\mathrm{res,t}$.
Concurrently, a Markov predictor\footnote{Note that this Markov predictor applies Markov chains on the \emph{delta} space and not the address space. It is essentially a delta correlating prefetcher designed for constant or alternating deltas only.} produces its own prediction
$\hat{y}_\mathrm{Markov,t}$.
A confidence gate selects the higher-confidence prediction for issuing prefetches,
making \ours~a \emph{tournament prefetcher}: the Markov predictor handles simple,
repetitive delta patterns, while the reservoir covers everything else.
A prefetch filter suppresses redundant requests and, by tracking the ratio of issued to
useful prefetches, dynamically throttles the prefetch lookahead degree.

A per-PC state stores the reservoir's metadata: the last address seen by $\mathrm{PC}_\mathrm{x}$ (to compute
$\mathtt{delta}_\mathtt{t}$), the PC's own weights $W_\mathrm{out}$, and the previous prediction $\hat{y}_\mathrm{res,t-1}$,
which is verified with $\mathtt{delta}_\mathtt{t}$. Together with $\mathtt{delta}_\mathtt{t}$ and the reservoir at $\mathtt{t-1}$, $\mathcal{O}_\mathrm{reservoir,t-1}$,
it drives a stochastic gradient descent~(SGD) update of $W_\mathrm{out}$ (SGD is more hardware-friendly than ridge regression).

For timeliness, the fill latency of each prefetched line is folded into a global exponentially weighted
moving average~(EWMA) of prefetch latency, and each demand access updates a per-PC EWMA of the
inter-access gap for $\mathrm{PC}_\mathrm{x,t}$. The ratio of prefetch latency to inter-access gap sets the
look-ahead degree, adapting it to both the prevailing memory latency and each PC's access rate
(Section~\ref{sec:timeliness}). The rest of this section describes how features and thus the reservoir are
generated after every demand access, how weights are trained online, and how delta predictions are made.

\begin{figure*}[htp]
    \centering
    \begin{subfigure}[b]{0.41\linewidth}
        \centering
        \includegraphics[width=\linewidth]{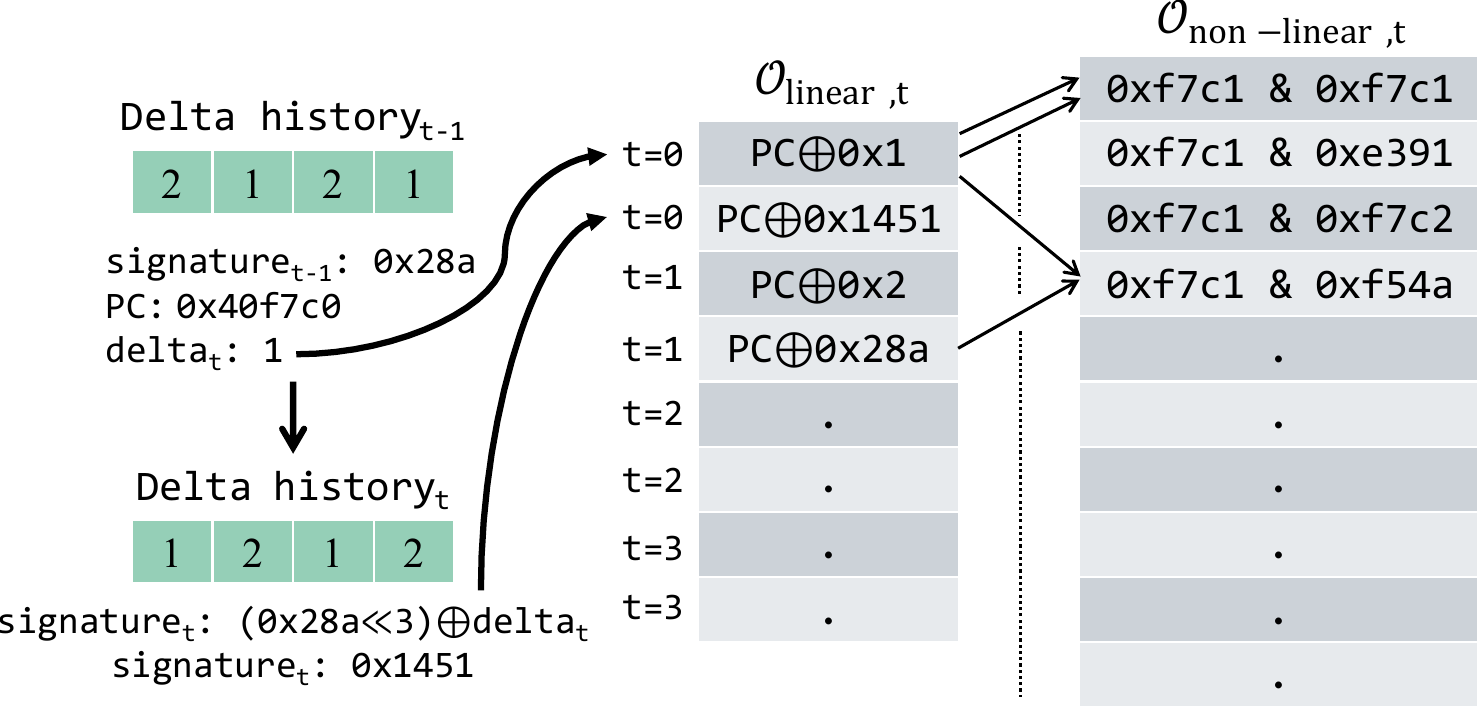}
        \caption{}
        \label{subfig:feature_gen}
    \end{subfigure}
    \hfill
    \begin{subfigure}[b]{0.57\linewidth}
        \centering
        \includegraphics[width=\linewidth]{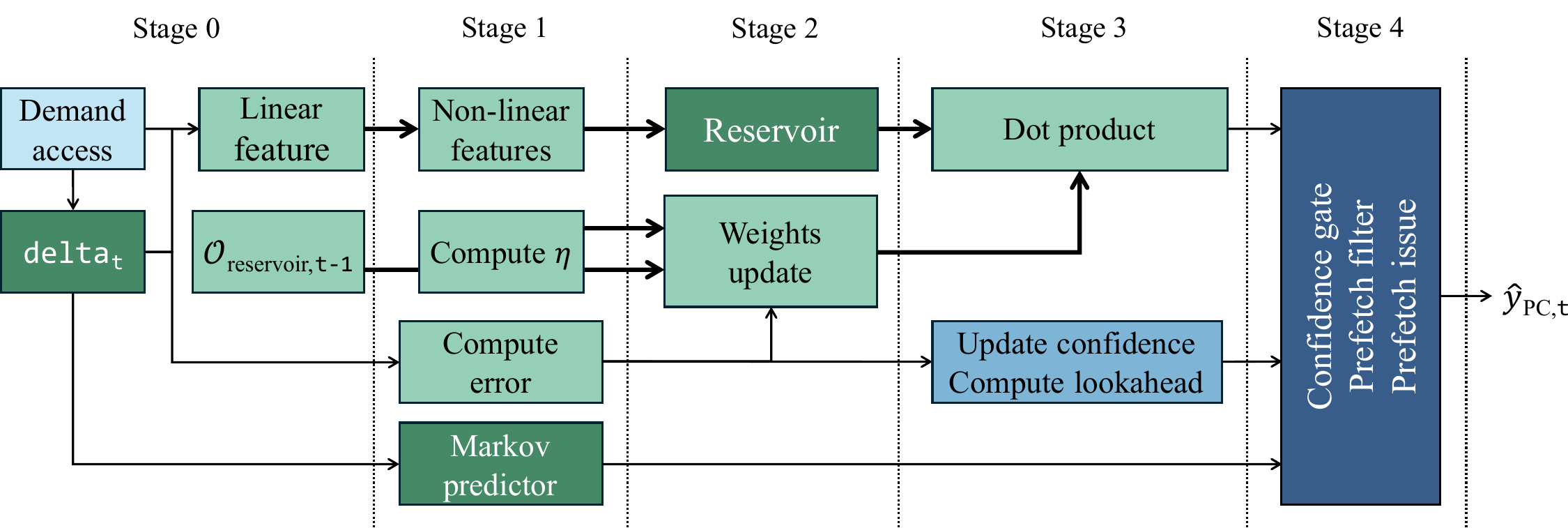}
        \caption{}
        \label{subfig:design_flow}
    \end{subfigure}  
    
    \caption{
    (a) Input features for \ours: $PC \oplus \mathtt{delta_t}$ and $PC \oplus \mathtt{signature_t}$ ($k=4$);
    (b) Pipelined view of \ours.
    }
    \vspace{-0.15in}
    \label{fig:design}
\end{figure*}

\subsection{Feature and reservoir generation}
Every demand access generates input features that are pushed into a context window of linear features, from which
non-linear features are generated; both are used for prediction. Figure~\ref{subfig:feature_gen} shows this flow, and each stage is explained below.

\subsubsection{Input features}
The \emph{quality} and \emph{quantity} of input features are crucial design choices: quality dictates how well the reservoir learns delta patterns,
while the number of input features $f$ determines the reservoir size. The number of linear features is $N = f \times k$,
where $k$ is the number of time-delayed history entries, and the number of non-linear features scales polynomially with $N$:
for $d=2$, it is $\binom{N+2-1}{2} = \frac{N^2 + N}{2}$ (including square terms), so the reservoir grows quadratically with $f$.

\begin{table}[t]
\centering
\caption{Features used for \ours~(assuming 64B lines).}
\label{tab:inp_features}
\resizebox{\columnwidth}{!}{%
\begin{tabular}{@{}ccc@{}}
\toprule
\textbf{Feature}   & \textbf{Computation}                            & \textbf{Feature width} \\ \midrule
$PC \oplus \mathtt{delta_t}$         & $PC \oplus ((Addr_\mathtt{t} \gg 6) - (Addr_{\mathtt{t-1}} \gg 6))$                         & 20 bits                \\
$PC \oplus \mathtt{signature_t}$      & $PC \oplus (\mathtt{signature_{t-1}} \ll 3) \oplus \mathtt{delta_t}$ & 20 bits                \\ \bottomrule
\end{tabular}%
}
\vspace{-0.1in}
\end{table}

Table~\ref{tab:inp_features} shows the two features we selected, both used by Pythia~\cite{pythia}, with the signature first introduced by SPP~\cite{spp}.
$PC \oplus \mathtt{delta_t}$ provides the instantaneous delta, while
$PC \oplus \mathtt{signature_t}$ provides a history of local (per-PC) deltas; hashing both with the PC
adds control flow information to the reservoir.
Because $PC \oplus \mathtt{signature_t}$ already encodes a rolling history of recent strides, the context window length $k$ can be kept small without sacrificing sequential context.
Two features also keep the reservoir from scaling unsustainably:
with $f=2$, $d=2$, and $k=16$, $N=32$, yielding just $\binom{32+2-1}{2} = 528$ non-linear features, whereas a third input feature would be costly ($f=3$, $d=2$, $k=16$, $\binom{48+2-1}{2} = 1176$).
All features are 20-bit integers, enough range to distinguish distinct stride
patterns while keeping the dot-product arithmetic within bounded integer precision.

\subsubsection{Generating linear features}
As shown in Figure~\ref{subfig:feature_gen}, the input features at time-step $\mathtt{t}$ (short-hand $\mathtt{a_t}$ and $\mathtt{b_t}$) are pushed into a fixed time-delayed
context window $\mathcal{O}_{\mathrm{linear,t}}$ of length $k$, holding features of the $k$ most recent demand accesses.
As shown in the original work on next-generation reservoir computing~\cite{Gauthier2021}, time-delayed inputs
let the output readout layer access trajectories of input features over time, not just instantaneous values,
so it can learn temporally encoded patterns.
We fix $k=16$, so the linear features have $N = f \times k = 2 \times 16 = 32$ elements, stored
in a time-major flattened register, i.e., the entries for a time step are always adjacent.
We term the combination of linear and non-linear features the reservoir for simplicity.
Until $\mathcal{O}_\mathrm{linear,t}$ has entries for $k$ time-steps, \ours~issues next-line prefetches (NLPs) into the last-level cache (LLC) to warm up.

\subsubsection{Generating non-linear features}
The original formulation of next-generation reservoir computing used multiplications between linear features to generate
non-linear features. 
For $f=2$, $d=2$, and $k=16$, $N=32$, generating non-linear features with multiplications would require $\binom{32+2-1}{2} = 528$
integer multipliers, a modest if not expensive hardware cost. 
Multiplications are used purely to introduce non-linearities into the reservoir. A key question is: 

\begin{center}
    \emph{Can we use more efficient bitwise operations for generating non-linear features instead of multiplications?}
\end{center}

As it turns out, \textbf{we can use bitwise AND operations between linear features instead of integer multiplications while still
exploiting the advantages of next-generation reservoir computing}. 
Figure~\ref{subfig:feature_gen} shows how non-linear features $\mathcal{O}_\mathrm{non-linear,t}$ at time-step $\mathtt{t}$ with $d=2$ are generated by taking pairs of linear features and
computing the bitwise AND between them. 

Our rationale for using bitwise-AND operations is as follows. For prefetching,
features are hash-derived, so what matters is which bits are set rather than
precise numeric values. This motivates replacing multiplication with bitwise AND.

To see why AND is sufficient, note that integer multiplication between two
$B$-bit integers $X$ and $Y$ expands as
$X \times Y = \sum_{i,j}(x_i \land y_j) \times 2^{i+j}$,
where $x_i$ and $y_j$ are individual bits. A bitwise AND retains exactly the
$B$ diagonal terms where $i = j$ i.e., the products of corresponding bits, while
discarding the $B^2 - B$ off-diagonal cross-terms. Though this is only $1/B$
of the full bit-product expansion, it is sufficient to ensure non-linear
coupling between features: each bit of $X \land Y$ is a conjunction of bits
from both $X$ and $Y$, a dependency no linear combination of $X$ and $Y$
could produce.

For conventional NGRC, features are typically real numbers from dynamical
systems, where multiplication is necessary since bitwise operations would
discard potentially useful information. Reservoir computing --- the precursor
to NGRC --- has variants built on boolean
functions~\cite{Canaday2018, Yilmaz2014, Kleyko2022}.
As a bonus, the $B$-bit output has no overflow or truncation, is computed in a single cycle, and is concatenated with
$\mathcal{O}_{\mathrm{linear,t}}$ to form the reservoir state $\mathcal{O}_{\mathrm{reservoir,t}}$, feeding the linear readout on equal footing with the linear features.
\rev{An AND feature needs 20 AND gates, a $20\times20$ array multiplier 400 AND gates and 380 adders;
the readout and the SGD update still multiply (Section~\ref{sec:inference}). The 528 include 32 square terms, kept for consistency with NGRC;
with AND they equal the linear features ($X \wedge X = X$), and since each square weight starts at zero and receives
the same updates as its linear twin, the two weights are always equal. \ours~therefore stores one weight per pair and doubles
the linear term with a 1-bit shift, which is bit-identical, so each PC stores 528 weights.}

\subsection{Online training and inference} \label{sec:iiid}
Using bitwise AND operations instead of integer multiplications \rev{for feature generation}
significantly reduces \ours{}'s hardware overhead.
Figure~\ref{subfig:design_flow} shows \ours's pipelined training and inference datapaths.
On every demand access, the delta $\mathtt{delta_t}$ of the current address from the PC's previously seen address is computed;
it is the ground truth of \ours's \emph{previous} prediction.

\subsubsection{Training}
$\mathtt{delta_t}$ gives the $error = \mathtt{delta_t}-\hat{y}_\mathrm{res, t-1}$ ($\hat{y}_\mathrm{res, t-1}$ is stored in the per-PC state),
which drives an SGD update of that PC's weights $W_\mathrm{out}$; every integer weight is clamped to 20 bits.
A zero error increments the demand PC's confidence; otherwise, it is decremented.
\ours~thus learns \textbf{online} after every demand access, making it robust across program phases since it
can easily learn new delta patterns.

\subheading{SGD updates}
From the error computed with $\mathtt{delta_t}$, the weights are updated with element-wise multiplications: 
\begin{equation}
W_\mathrm{out} \leftarrow W_\mathrm{out} + \eta \odot error \odot \mathcal{O}_\mathrm{reservoir, t-1}
\end{equation}

Here $\eta$ is the learning rate and $\mathcal{O}_\mathrm{reservoir, t-1}$ is the reservoir (linear and non-linear features) used for generating $\hat{y}_\mathrm{res, t-1}$. 
Rather than storing the entire reservoir per PC, we regenerate $\mathcal{O}_\mathrm{reservoir, t-1}$ from the stored $\mathcal{O}_\mathrm{linear, t-1}$,
since non-linear feature generation is extremely cheap. This is done in the first datapath stage (Figure~\ref{subfig:design_flow}),
time-multiplexing the hardware generating the non-linear features for $\mathcal{O}_\mathrm{reservoir, t}$.
SGD, rather than ridge regression~\cite{Gauthier2021}, also reduces hardware overhead:
ridge regression updates weights as $W_\mathrm{out} \leftarrow \lambda \odot \mathrm{W_{decay}} \odot W_\mathrm{out} + \eta \odot error \odot \mathcal{O}_\mathrm{reservoir, t-1}$, requiring extra multiplications for the L2 regularization (decay).

\subheading{Adaptive learning rate}
Rather than a fixed learning rate, we use an adaptive schedule driven by the exponentially weighted moving average of the per-PC prediction residual magnitude,
\begin{equation}
  \bar{e} \;\leftarrow\; \bar{e} - (\bar{e} \gg 5) + (|\mathtt{delta_t} - \hat{y}_\mathrm{res,t-1}| \gg 5)
\end{equation}
giving the learning rate --
\begin{align*}
  s = [\bar{e} \geq 1] + [\bar{e} \geq 3] + [\bar{e} \geq 7] + [\bar{e} \geq 15];\eta = 2^{-(12+s)}
\end{align*}
When the PC is in a volatile phase with large residuals, $\eta$ is reduced, preventing weight thrashing. During stable phases the rate
rises back to $2^{-12}$, allowing faster adaptation. Note that this is the \emph{opposite} of standard adaptive optimizers
such as Adam, which increase the per-parameter rate when gradients are large; here the goal is stability over speed.

\subsubsection{Inference}\label{sec:inference}
\sloppy
Once $\mathcal{O}_{\mathrm{linear,t}}$ and $\mathcal{O}_{\mathrm{non-linear,t}}$ form the current reservoir state $\mathcal{O}_{\mathrm{reservoir,t}}$, $\hat{y}_\mathrm{res,t}$ is a simple dot-product with the updated weights $W_\mathrm{out}$,
requiring integer multiplications between features and weights and integer adders for the reduction.
Though Figure~\ref{subfig:design_flow} shows the dot-product in a single stage, it can be split over multiple
clock cycles (especially the reduction).
\rev{The Q20 dot product is shifted right by 40 bits and the per-PC bias (clipped to $\pm(2^{20}-1)$, so
it contributes at most one line) is added; candidates outside the trigger's 4~KB page are dropped, so
the output is a 7-bit saturating delta with a valid bit. Training clips the error to $[-128,127]$ and
skips deltas with $|delta_t| \geq 64$; deltas beyond $\pm128$ skip the reservoir and enter the Markov context as a
reserved token. Features use the low 20 bits of the PC. A back-to-back same-PC access predicts with the
not-yet-updated weights (delayed SGD, no stall); we evaluate the effect of this delay in Section~\ref{sec:learning}.}

\subheading{Regression vs. classification}
\ours~treats delta prediction as regression rather than classification, giving it a dynamic range of $[-64,63]$ on physical addresses.
In contrast, other ML-based prefetchers such as Pythia can only choose deltas from their
action set, $\{-6,-3,-1,0,1,3,4,5,10,11,12,22,23,30,32\}$~\cite{pythia}, configured by offline optimization passes
on a fixed set of workloads. \rev{In our experiments, restricting \ours~to Pythia's action set does not hurt performance (it even improves GAP),
so the benefit of regression is that no action set has to be chosen offline, not the wider range.
The readout does not interpolate between hash values: weights are per PC and the hashed contexts of a PC recur exactly,
so it effectively learns which recurring contexts precede which delta, while multimodal patterns such as alternating
deltas are left to the Markov table (Section~\ref{sec:markov}).}

\subsubsection{Per-PC state}
\sloppy
Table~\ref{tab:pcstate} lists each entry of the per-PC state (Section~\ref{sec:overview}), its function, and its storage.
$W_\mathrm{out}$ is the greatest part of this storage, so to balance the number of tracked PCs ($N_{\mathrm{PC}}$) against total storage, we fix $N_{\mathrm{PC}}=16$;
Section~\ref{sec:eval} studies the sensitivity of \ours's performance to $N_{\mathrm{PC}}$. Per-PC states are kept in a fully-associative structure, evicted by LRU time step when full.

\begin{table}[h!]
\centering
\vspace{-0.1in}
\caption{Entries and storage required for the per-PC state.}
\label{tab:pcstate}
\resizebox{\columnwidth}{!}{%
\begin{tabular}{@{}llc@{}}
\toprule
\multicolumn{1}{c}{\textbf{Entry}} & \multicolumn{1}{c}{\textbf{Function}}   & \textbf{Storage required}                                          \\ \midrule
$W_\mathrm{out}$                   & Per-PC weights                          & \rev{$(N + \binom{N}{d})\times 20 = 528 \times 20 = 10560\mathrm{~bits}$} \\
Bias                               & Per-PC bias                             & 20 bits                                                            \\
LRU time-step                      & For evicting PC state entry             & 32 bits                                                            \\
$\hat{y}_\mathrm{res,t-1}$         & Last prediction for this PC             & 7 bits                                                             \\
$\mathcal{O}_\mathrm{linear, t-1}$ & For generating $\mathcal{O}_\mathrm{reservoir, t-1}$ to update $W_\mathrm{out}$           & $N \times 20 = 32 \times 20 = 640\mathrm{~bits}$                    \\
$c_{reservoir}$                    & Confidence for this PC                  & 3 bits                                                             \\
$\bar{e}$                          & EWMA error tracker                      & 16 bits                                                             \\
Last access cycle                  & Last cycle this PC was accessed         & 16 bits                                                            \\
Average access gap                 & Average gap in cycles between accesses  & 16 bits                                                            \\
Delta history                      & Last 8 deltas for this PC               & $8 \times 7 = 56\mathrm{~bits}$                                     \\
Valid                              & To initialize this PC state             & 1 bit                \\ \hline  
\multicolumn{2}{l}{\textbf{Total}}                                           & \rev{1.39 KB per PC} \\ \bottomrule
\end{tabular}%
}
\vspace{-0.2in}
\end{table}

\subsection{Markov table for trivial predictions}\label{sec:markov}

Many benchmarks have phases in which PCs have trivial delta patterns such as constant or alternating deltas, e.g.,
a steady delta of one for PC \texttt{0x40f7c0} in \texttt{602.gcc}, or deltas alternating between two and three for PC \texttt{0x401cb0} in \texttt{619.lbm}.
Here the recent delta context can entirely determine the next delta, so a lightweight lookup table suffices:
the Markov predictor~\cite{markov} maps a \emph{context} tuple to a predicted next delta, a simple, low-cost path that bypasses the reservoir.

\subsubsection{Context and table organization}
\sloppy
The context for a given load $PC_\mathrm{t}$ at time step $\mathtt{t}$ is the pair of the two most recently observed cache-line deltas,
$(\mathtt{delta}_{\mathtt{t-1}},\, \mathtt{delta}_{\mathtt{t-2}})$. The context $(PC_{\mathrm{t}},\, \mathtt{delta}_{\mathtt{t-1}},\, \mathtt{delta}_{\mathtt{t-2}})$ is hashed to produce both a \emph{table index} $\iota$ and a partial
\emph{tag} $\tau$. The global Markov table contains $N{=}256$ direct-mapped entries; each entry stores a tag $\tau$, the next delta $\hat{y}_{\mathrm{Markov,t}} \in [-64,63]$, the previous two deltas, a 3-bit saturating confidence counter $c_{\mathrm{Markov}} \in [0,7]$, and a valid bit.

\subsubsection{Update rule}

On every memory access the table entry at index $k$ is updated as follows. $\mathtt{delta}_{\mathrm{t}}$ is the observed delta.

\begin{enumerate}
  \item \textbf{Hit} ($\mathtt{valid}=1$ and $\tau$ matches): overwrite
        $\hat{y}_{\mathrm{Markov,t}} \leftarrow \mathtt{delta}_{\mathtt{t}}$ and saturate $c_{\mathrm{Markov}} \leftarrow 7$.
  \item \textbf{Conflict} ($\mathtt{valid}=1$ and $\tau$ mismatches):
        if $c_{\mathrm{Markov}} \leq 2$, evict and install the new entry with $c_{\mathrm{Markov}}
        \leftarrow 4$; otherwise decrement $c_{\mathrm{Markov}} \leftarrow c_{\mathrm{Markov}} - 1$.
  \item \textbf{Empty} ($\mathtt{valid}=0$): install $(\tau, \mathtt{delta}_\mathrm{t}, c_{\mathrm{Markov}}{=}7)$.
\end{enumerate}

During the update, the history registers are shifted: $\mathtt{delta}_{\mathtt{t-2}} \leftarrow \mathtt{delta}_{\mathtt{t-1}}$; $\mathtt{delta}_{\mathtt{t-1}} \leftarrow \mathtt{delta}_\mathtt{t}$.

\subsubsection{Prediction}
\sloppy
The \emph{next} delta is predicted by immediately looking up the now-updated context $(PC_\mathrm{t},\, \mathtt{delta}_\mathrm{t},\, \mathtt{delta}_{\mathrm{t-1}})$, valid only when the tag matches and $c_{\mathrm{Markov}} \geq 2$.
Per-access feedback increments $c_{\mathrm{Markov}}$ on a correct prediction and decrements it otherwise, a second confidence loop independent of the table update.

\subsection{Final prefetch candidate}

The final prefetch candidate $\hat{y}_\mathtt{PC,t}$ (subject to filtering) is dependent on the confidence of both the reservoir and Markov predictions.
We use a simple heuristic for determining the source of the prefetch: 
\begin{equation*}
    \mathrm{pref.~candidate} =
    \begin{cases}
        \hat{y}_\mathrm{Markov,t}               & c_\mathrm{Markov} \geq c_{\mathrm{reservoir}} + 4, \\
        \mathrm{next-line}            & c_{\mathrm{reservoir}} < 1, \\
        \hat{y}_\mathrm{res,t}            & \mathrm{otherwise}.
    \end{cases}
    \label{eq:issuance}
\end{equation*}
The Markov predictor takes priority over the reservoir only when it is significantly more confident ($c_{\mathrm{Markov}} \geq c_{\mathrm{reservoir}} + 4$).
When the reservoir's confidence is zero, \ours~falls back to simple NLPs into the LLC.
A confidence threshold of one is aggressive for issuing prefetches; the next section shows how
useless next-line prefetches are throttled.

%% file: inputs/5Design2.tex

\section{Prefetch Filtering and Timeliness}
\label{sec:timeliness}

\sloppy
Filtering redundant/useless prefetches and ensuring timely prefetches~\cite{navarro2022berti, fdp, ppf} is 
crucial for improving the performance of any prefetcher. We now define these mechanisms
for \ours. \rev{Unlike prior usefulness-based throttling~\cite{fdp}, \ours~tracks usefulness per PC and separately for its predictors and its next-line fallback, and none of these mechanisms depend on the predictor, so they can be used with other prefetchers as well (Section~\ref{sec:sources}).}

\begin{figure}[t]
    \centering
    \begin{subfigure}[b]{0.63\linewidth}
        \centering
        \includegraphics[width=\linewidth]{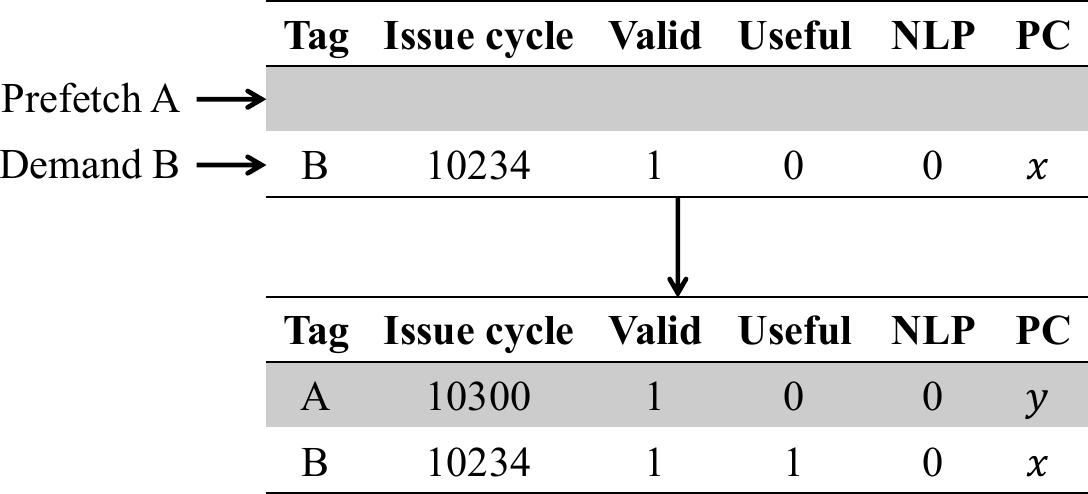}
        \caption{}
        \label{subfig:filter}
    \end{subfigure}
    \hfill
    \begin{subfigure}[b]{0.35\linewidth}
        \centering
        \includegraphics[width=\linewidth]{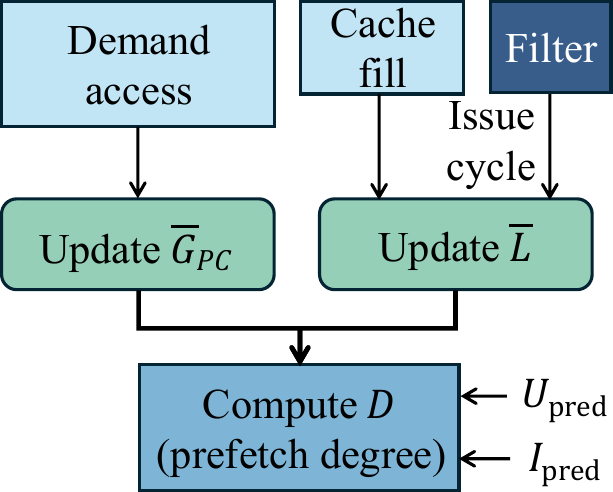}
        \caption{}
        \label{subfig:lookahead}
    \end{subfigure}  
    
    \caption{
    (a) Prefetch filter with fields for tracking NLPs and the prefetch issue cycle;
    (b) Computing the prefetch degree.
    }
    \label{fig:timeliness}
    \vspace{-0.15in}
\end{figure}

\subsection{Filtering prefetch requests}

\ours~filters prefetches at two levels. First, two independent arrays of 1024 9-bit saturating-counter pairs, indexed by a 10-bit load-PC hash, track \emph{issued} and \emph{useful} counts for \ours~prefetches ($I_{\mathrm{pred}}, U_{\mathrm{pred}}$) and fallback next-line prefetches ($I_{\mathrm{nlp}}, U_{\mathrm{nlp}}$) \emph{per-PC}, throttling/disabling prefetching for a PC with low usefulness. Next-line prefetches get separate counters to prevent cache pollution when many PCs are unpredictable (e.g., linked-list traversals in \texttt{mcf}).
Both counter pairs are halved once either issued count reaches 512, preserving sensitivity to recent behavior without indefinitely penalizing a PC.

At the second level, \ours~employs a prefetch filter (Figure~\ref{subfig:filter}) modeled after SPP~\cite{spp}, extended with per-PC usefulness tracking to drive throttling decisions. The filter is direct-mapped with $N_F = 1024$ sets. 
Each entry stores a tag derived from a Wang hash of the cache-line address, a \texttt{valid} bit, a \texttt{useful} bit, the issue cycle, a flag distinguishing predicted prefetches from next-line prefetches, and a 10-bit PC hash for crediting usefulness trackers.


After usefulness-based filtering, a candidate prefetch is admitted only if its hashed line address maps to an empty or tag-mismatching slot in the filter, preventing the prefetch queue (PQ) from being flooded with recently prefetched or cache-resident lines. 
On a demand access, the filter marks the matching entry as useful and credits the prediction tracker if the line was issued by \ours. \rev{A prefetch is \emph{useful} if its line is demanded before eviction and \emph{useless} otherwise;
filtering is one direct-mapped lookup per candidate, off the demand path.}
When a line is evicted from the cache, the corresponding filter slot is invalidated, freeing it for future prefetches.






\subsection{Timeliness and lookahead}

Issuing a fixed number of prefetches regardless of memory access rate risks either 
under-prefetching (cache misses still prevalent resulting in stalls) or over-prefetching (cache and bandwidth 
pollution). \ours~estimates a \emph{timely lookahead degree} at runtime by 
maintaining two exponentially weighted moving averages (EWMA):

\begin{itemize}
    \item $\overline{G}_{\mathrm{PC}}$: per-PC average cycles between
    successive accesses by the same load instruction.
    \item $\overline{L}$: global average prefetch fill latency, i.e., cycles from
    placing a prefetch in the filter to its fill.
\end{itemize}

\noindent Both are tracked in 16-bit registers. Figure~\ref{subfig:lookahead} shows how these approximate latencies yield
the lookahead, keeping prefetches within a timely window.
Once a PC and the global latency estimate have at least 8 latency samples, the minimum degree covering memory latency is--

\begin{equation}
    D_{\mathrm{timely}} = 
    \left\lceil \frac{\overline{L}}{\overline{G}_{\mathrm{PC}}} \right\rceil
    \label{eq:timely_degree}
\end{equation}

\noindent capped at $D_{\max} = 2 \times D_{\mathrm{base}}$, where 
$D_{\mathrm{base}} = 6$ is the nominal prefetch degree (similar to~\cite{pythia}). The lookahead degree $D_{\mathrm{timely}}$
avoids explicit division in hardware by inverting the comparison: for each candidate $D_\mathrm{cand.} \in \{1,\ldots,D_{\max}\}$, check whether
$D_\mathrm{cand.} \cdot \overline{G}_{\mathrm{PC}} \geq \overline{L}$; the minimum passing $D_\mathrm{cand.}$ is selected via a priority encoder.


\begin{figure}[t]
    \centering
    \includegraphics[width=\linewidth]{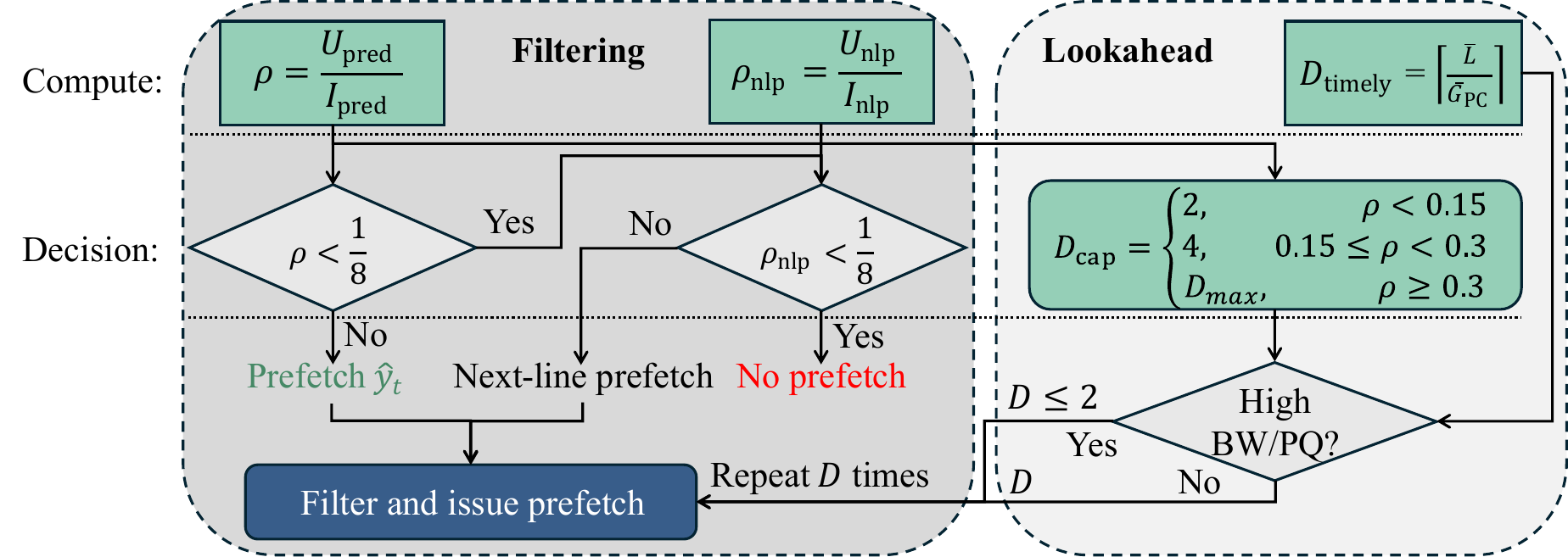}
    \caption{Overall flow for prefetch issuance.}
    \label{fig:timeliness_flow}
    \vspace{-0.1in}
\end{figure}

\subsection{Putting it all together}
Figure~\ref{fig:timeliness_flow} shows the overall prefetch issuance flow once the reservoir makes a prediction. 
First, the usefulness ratios and $D_{\mathrm{timely}}$ are computed. Then, the usefulness ratios are used to filter predictions, and in case
they are extremely low, issue no prefetches at all. We use \rev{a ratio of 1/8 (a 12.5\% usefulness rate)} for triggering the filter. 
Concurrent to the filtering steps, the lookahead degree $D_{\mathrm{cap}}$ is determined using the prediction usefulness ratio $\rho$. 
These values of $D_\mathrm{cap}$ were fixed and not tuned for different benchmarks.
However, the intuition behind these values is that when the usefulness of prefetching is low for a PC, 
deep lookahead prefetches will have limited impact for that PC. \rev{These ratio comparisons need no division: multiplying both sides by $I_{pred}$ turns them into
integer shifts and compares.}
 Using
$D_{\mathrm{timely}}$ and $D_{\mathrm{cap}}$, we determine the final prefetch degree $D$: 

\begin{equation}
    D = \min\!\left(\min\!\left(D_{\mathrm{timely}},\, D_{\max}\right),\,D_{\mathrm{cap}}\right)
    \label{eq:final_degree}
\end{equation}



An additional hard cap of $\min\left(D, 2\right) \implies D \leq 2$ applies when PQ occupancy exceeds 90\% or DRAM bandwidth utilization crosses a high-bandwidth threshold, so the prefetcher does not stress the DRAM interface even when highly confident.

\ours~generates the $D$ candidate addresses by linear delta extrapolation from the current address;
Markov predictions instead walk the Markov transition table
$D$ steps from the current delta-history context, stopping early if any
intermediate entry is invalid or predicts a zero delta.

\subsection{Total storage cost}

Table~\ref{tab:storage} shows the storage of \ours, with all weights and features fixed as 20-bit integers.
\ours~requires slightly more storage than prefetchers like Pythia~\cite{pythia}, still $\sim3\%$ of
a moderately sized 1 MB L2 cache; per-PC states constitute two-thirds of the total.

\begin{table}[h!]
\centering
\caption{Storage overhead of \ours.}
\label{tab:storage}
\resizebox{\columnwidth}{!}{%
\begin{tabular}{@{}llc@{}}
\toprule
\multicolumn{1}{c}{\textbf{Component}} & \multicolumn{1}{c}{\textbf{Description}} & \textbf{Storage}               \\ \midrule
$\mathcal{O}_{\mathrm{linear,t}}$      & Global context window                    & 640 bits $\rightarrow$ 0.08 KB \\
PC states $\times N_\mathrm{PC}$       & All per-PC states                        & \rev{1.39 KB $\times$ 16 = 22.2 KB} \\
Global latency tracking                & For timely prefetches                    & 16 bits $\rightarrow$ 0.002 KB \\
Prefetch filter                        & Filtering \ours~prefetches               & 4.87 KB                        \\
Markov table                           & 256-entry, 49 bits / entry               & 1.53 KB                        \\
Prefetch usefulness trackers           & $U_\mathrm{pred}, I_\mathrm{pred}$ etc.  & 4.5 KB                         \\ \midrule
\multicolumn{2}{l}{\textbf{Total}}                                                & \rev{33.18 KB}                       \\ \bottomrule
\end{tabular}%
}
\end{table}

%% file: inputs/6Methodology.tex
\section{Methodology}

{\noindent \textbf{Simulator configuration}} We evaluate \ours~with the trace-driven ChampSim simulator~\cite{champsim}, using the DPC-4 infrastructure~\cite{dpc4}, which contains the latest implementations of all evaluated prefetchers.
Table~\ref{tab:simulator} shows the configuration for each core count, with 4 KB pages and 64B cache lines.

\begin{table}[ht!]
\centering
\vspace{-0.1in}
\caption{Simulator configuration.}
\label{tab:simulator}
\resizebox{\columnwidth}{!}{%
\begin{tabular}{@{}cclll@{}}
\toprule
\textbf{Core}             & \multicolumn{4}{c}{1/4/8 6-wide OoO cores; 512-entry ROB; 128/72-entry LQ/SQ;}                                                                                                                                                                                                                   \\ \midrule
\textbf{Branch predictor} & \multicolumn{4}{c}{Perceptron~\cite{perceptron}; 17-cycle mispredict penalty}                                                                                                                                                                                                                                               \\ \midrule
\textbf{L1D/L2C}          & \multicolumn{4}{c}{\begin{tabular}[c]{@{}c@{}}Private; 48KB/1MB with 64B lines; 12/16-way; LRU replacement; \\ 16/48 MSHRs; 5-/15-cycle round-trip latency\end{tabular}}                                                                                                                                  \\ \midrule
\textbf{LLC}              & \multicolumn{4}{c}{\begin{tabular}[c]{@{}c@{}}Shared; 3MB/core; 64B lines, 12-way; SHiP~\cite{ship} \rev{(1C) / DRRIP (4C/8C)} replacement;\\ 64 MSHRs/bank; 55-cycle round-trip latency\end{tabular}}                                                                                                                                           \\ \midrule
\textbf{Main memory}      & \multicolumn{4}{c}{\begin{tabular}[c]{@{}c@{}}64b data bus; 2KB row buffer/bank;\\ \textbf{1C:} \rev{3200 MT/s;} single-channel, 1 rank/channel\rev{, 8 banks/rank;}\\ \textbf{4C:} \rev{4800 MT/s;} dual-channel, 2 ranks/channel\rev{, 16 banks/rank;}\\ \textbf{8C:} \rev{4800 MT/s;} quad-channel, 4 ranks/channel\rev{, 32 banks/rank;}\\ \rev{tRCD = tCAS = tRP = 12 (1C) / 36 (4C/8C) DRAM cycles (7.5 / 15 ns)}\end{tabular}} \\ \bottomrule
\end{tabular}%
}
\vspace{-0.12in}
\end{table}

{\noindent \textbf{Workloads}} We use the GAP~\cite{gap}, Ligra~\cite{ligra}, PARSEC~\cite{parsec}, SPEC 2006~\cite{spec2006}, and SPEC 2017~\cite{spec2017} suites (Table~\ref{tab:benchmarks} lists trace counts);
the GAP, Ligra, SPEC 2006, and SPEC 2017 traces are from the ML-DPC competition~\cite{mldpc}.
We use 137 traces across the five suites \rev{and 32 AI traces, 169 in total}.
SPEC06 and SPEC17 traces use 100M warmup and 500M simulated instructions;
all other suites \rev{use} 100M warmup and 150M simulated instructions.
For multi-core evaluations, we created 20 random heterogeneous benchmark mixes per suite (10M warmup + 50M simulated instructions).

\begin{table}[ht!]
\centering
\vspace{-0.1in}
\caption{All benchmark suites evaluated.}
\label{tab:benchmarks}
\resizebox{\columnwidth}{!}{%
\begin{tabular}{@{}ccc@{}}
\toprule
\textbf{Suite} & \textbf{Num. traces} & \textbf{Benchmarks}                                                                                                                                             \\ \midrule
GAP            & 16                   & \textit{bc, cc, pr, sssp}                                                                                                                                       \\
Ligra          & 17                   & \textit{\begin{tabular}[c]{@{}c@{}}BellmanFord, BFSCC, CF, Components, MIS, \\ PageRankDelta, PageRank, Radii, Triangle\end{tabular}}                           \\
PARSEC         & 25                   & \textit{\begin{tabular}[c]{@{}c@{}}blackscholes, bodytrack, canneal, dedup, facesim,\\ fluidanimate, raytrace, streamcluster, swapoptions,\\ vips\end{tabular}} \\
SPEC06         & 32                   & \textit{\begin{tabular}[c]{@{}c@{}}bwaves, mcf, milc, leslie3d, soplex, GemsFDTD,\\ libquantum, lbm, omnetpp, astar, sphinx\end{tabular}}                       \\
SPEC17         & 47                   & \textit{\begin{tabular}[c]{@{}c@{}}gcc, mcf, cactuBSSN, lbm, omnetpp, wrf,\\ xalancbmk, fotonik3d, roms\end{tabular}}                                           \\ \bottomrule
\end{tabular}%
}
\vspace{-0.1in}
\end{table}

{\noindent \textbf{Evaluated prefetchers}} Pythia~\cite{pythia} is our primary baseline as it outperforms prior delta prefetchers\rev{, including SPP with the PPF filter,} and is the state-of-the-art ML-based design. We also evaluate SMS~\cite{sms}, Berti~\cite{navarro2022berti}, and IPCP~\cite{ipcp}, absent from Pythia’s evaluation. IPCP and Berti operate on virtual addresses at L1D and may prefetch into any cache level; all other prefetchers operate on L1D misses and prefetch into the L2C/LLC. Table~\ref{tab:prefetchers} summarizes configurations and storage overheads. We exclude temporal prefetchers such as Twilight~\cite{duong2024new} as orthogonal to delta prefetching.
\rev{As in DPC-4, no prefetcher's decision latency is modeled and updates take effect immediately; a
prefetch is \emph{late} if a demand merges with it in flight.}

\begin{table}[ht!]
\centering
\vspace{-0.1in}
\caption{All evaluated prefetchers.}
\label{tab:prefetchers}
\resizebox{\columnwidth}{!}{%
\begin{tabular}{@{}lcc@{}}
\toprule
\textbf{Prefetcher} & \textbf{Configs}                                                                                                                                                              & \textbf{Storage overhead} \\ \midrule
Pythia              & \rev{unmodified DPC-4 port of}~\cite{pythia}                                                                                                                                                         & 25.5 KB                   \\
SMS                 & 2048-entry PHT; 256-entry prefetch buffer                                                                                                                                     & 22 KB                     \\
Berti               & \begin{tabular}[c]{@{}c@{}}64-entry current pages table; 1024-entry previous requests table;\\ 512-entry previous prefetches table; 1407-entry page record table\end{tabular} & 27.8 KB                   \\
IPCP                & 1024-entry IP table; 16-entry GHB; 6-bit tag                                                                                                                                  & 4.5 KB                    \\
\ours & \begin{tabular}[c]{@{}c@{}}16-entry PC states; 1024-entry prefetch filter; \\ 256-entry Markov table; 16-entry context window \end{tabular} & \rev{33.18 KB} \\ \bottomrule
\end{tabular}%
}
\vspace{-0.15in}
\end{table}

%% file: inputs/7Results.tex
\section{Results}
\label{sec:eval}

\begin{figure*}[htp]
    \centering
    \begin{subfigure}[b]{0.32\linewidth}
        \centering
        \includegraphics[width=\linewidth]{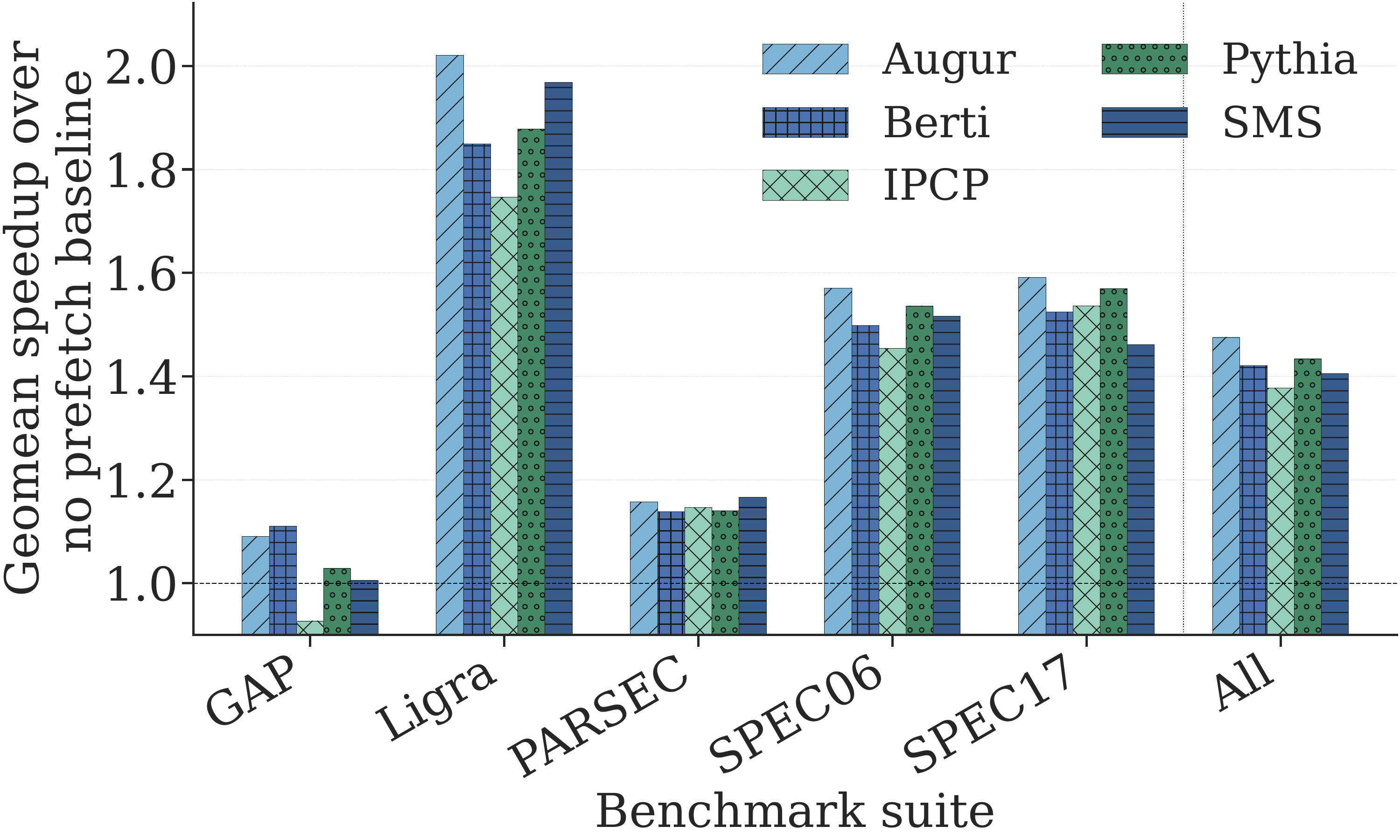}
        \caption{Single core.}
        \label{subfig:1c}
    \end{subfigure}
    \hfill
    \begin{subfigure}[b]{0.32\linewidth}
        \centering
        \includegraphics[width=\linewidth]{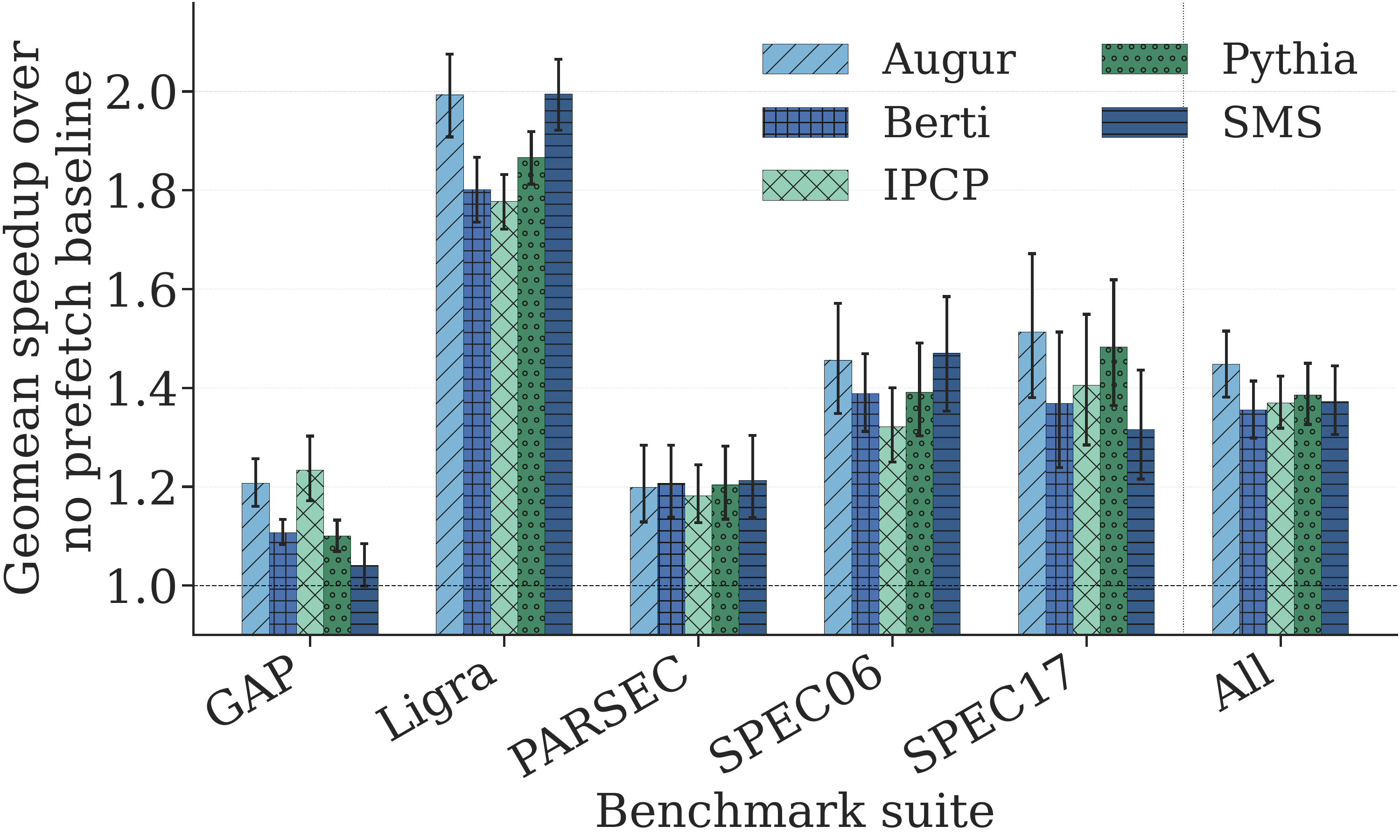}
        \caption{Four cores.}
        \label{subfig:4c}
    \end{subfigure}
    \hfill
    \begin{subfigure}[b]{0.32\linewidth}
        \centering
        \includegraphics[width=\linewidth]{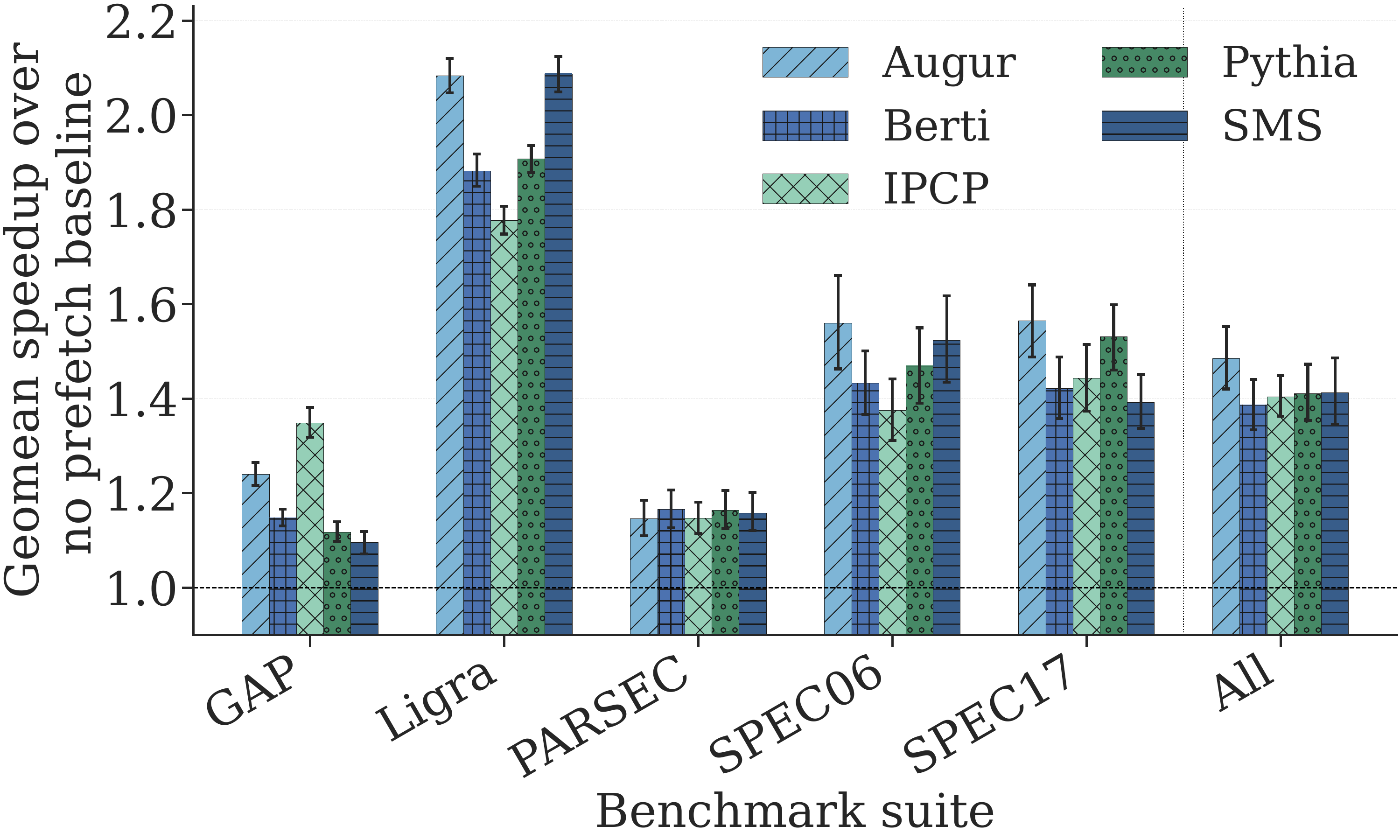}
        \caption{Eight cores.}
        \label{subfig:8c}
    \end{subfigure}
    \caption{
    (a)-(c) Speedups attained by using different prefetchers over a baseline with no prefetchers for different benchmark suites and core configurations. \rev{Error bars in (b) and (c) show 95\% bootstrap confidence intervals of the geomean over the 20 mixes per suite}.
     }
    \label{fig:speedups}
    \vspace{-0.1in}
\end{figure*}


\subsection{Speedup across benchmark suites}
Figure~\ref{fig:speedups} shows the speedups ($=\frac{IPC_\mathrm{prefetch}}{IPC_\mathrm{no\,prefetch}}$\rev{, with IPC summed over cores for multi-core runs}) of five prefetchers, including \ours, across benchmark suites and core configurations.

\paragraph{Single core simulations}
Figure~\ref{subfig:1c} shows the single core simulation results.
Overall, \ours~achieves a relative speedup of 3.85\%, 7.13\%, 2.9\% \rev{(95\% bootstrap CI [1.0, 4.8]\%)}, and 4.9\% over Berti, IPCP, Pythia, and SMS respectively. 
Notably, \textbf{\ours~outperforms Pythia over all five suites} GAP, Ligra, PARSEC, SPEC 2006, and SPEC 2017 by 6.0\%, 7.61\%, 1.52\%, 2.28\%, and 1.37\% respectively.
Berti outperforms \ours~on GAP by 1.78\%, as GAP's irregular access patterns are more amenable to L1D prefetchers that operate on virtual addresses.


\paragraph{Four core simulations}
With four cores (Figure~\ref{subfig:4c}), \ours~achieves a relative speedup of 6.8\%, 5.73\%, 4.5\%, and 5.48\% over Berti, IPCP, Pythia, and SMS respectively.
Pythia outperforms \ours~only on PARSEC (by 0.4\%), since \ours~prefetches more selectively than Pythia;
\ours~outperforms Pythia by 6.79\%, 9.75\%, 4.63\%, and 2.08\% on Ligra, GAP, SPEC 2006, and SPEC 2017.
\rev{IPCP outperforms \ours~on GAP by 2.1\%.}

\paragraph{Eight core simulations}
With eight cores (Figure~\ref{subfig:8c}), \ours~achieves a relative speedup of 7.12\%, 5.78\%, 5.3\%, and 5.16\% over Berti, IPCP, Pythia, and SMS respectively. As with four cores, IPCP outperforms \ours~on GAP, but \ours~outperforms Berti, Pythia, and SMS there by 8.04\%, 10.93\%, and 13.18\%, respectively.

\subheading{Summary} 
Across the five suites, \ours~\rev{achieves the highest geomean speedup} on single- and multi-core systems \rev{(Berti and IPCP lead on GAP)}, including Pythia by 2.9\%, 4.5\%, and 5.3\% with 1, 4, and 8 cores, respectively.

\begin{figure}[t]
    \centering
    \begin{subfigure}[b]{0.85\linewidth}
        \centering
        \includegraphics[width=\linewidth]{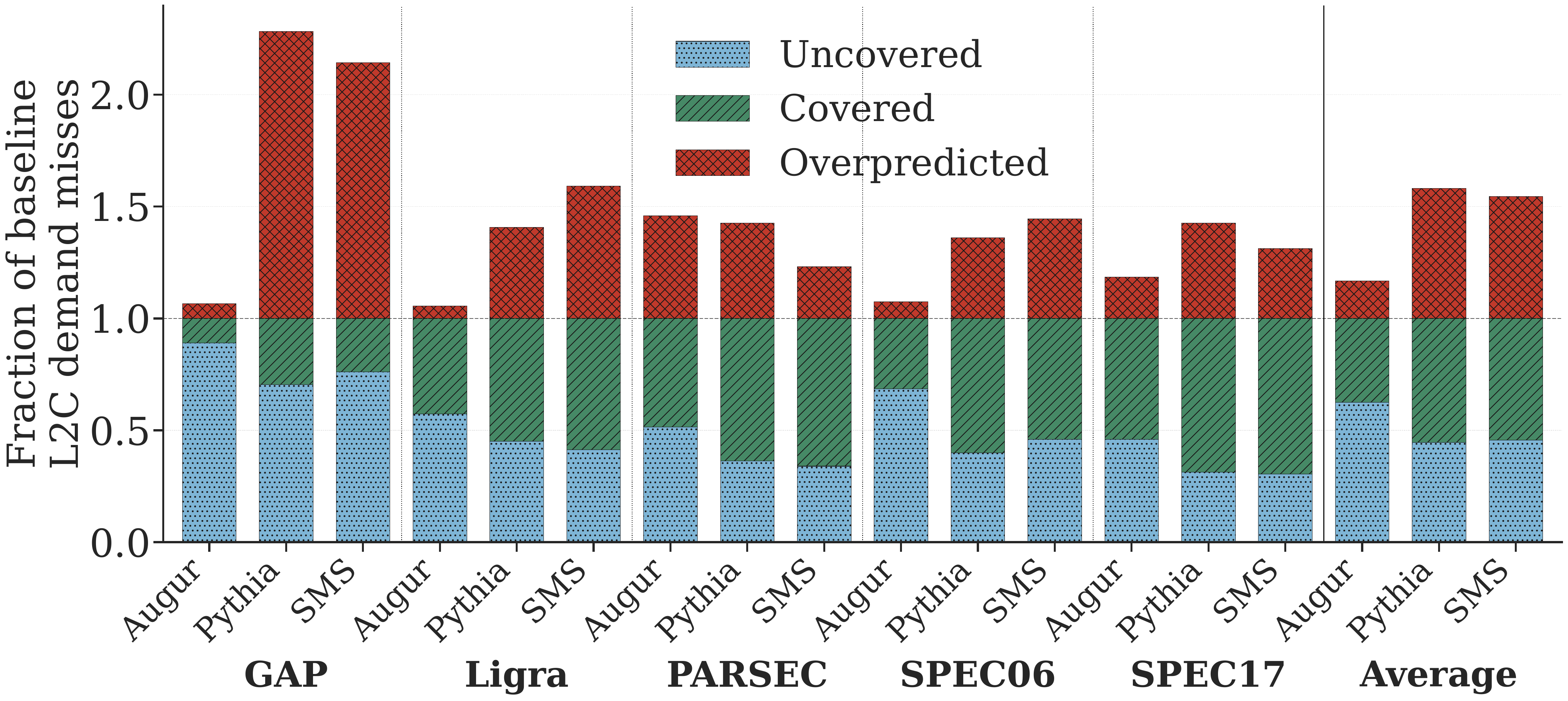}
        \caption{}
        \label{subfig:coverage}
    \end{subfigure}  
    \\
    \begin{subfigure}[b]{0.85\linewidth}
        \centering
        \includegraphics[width=\linewidth]{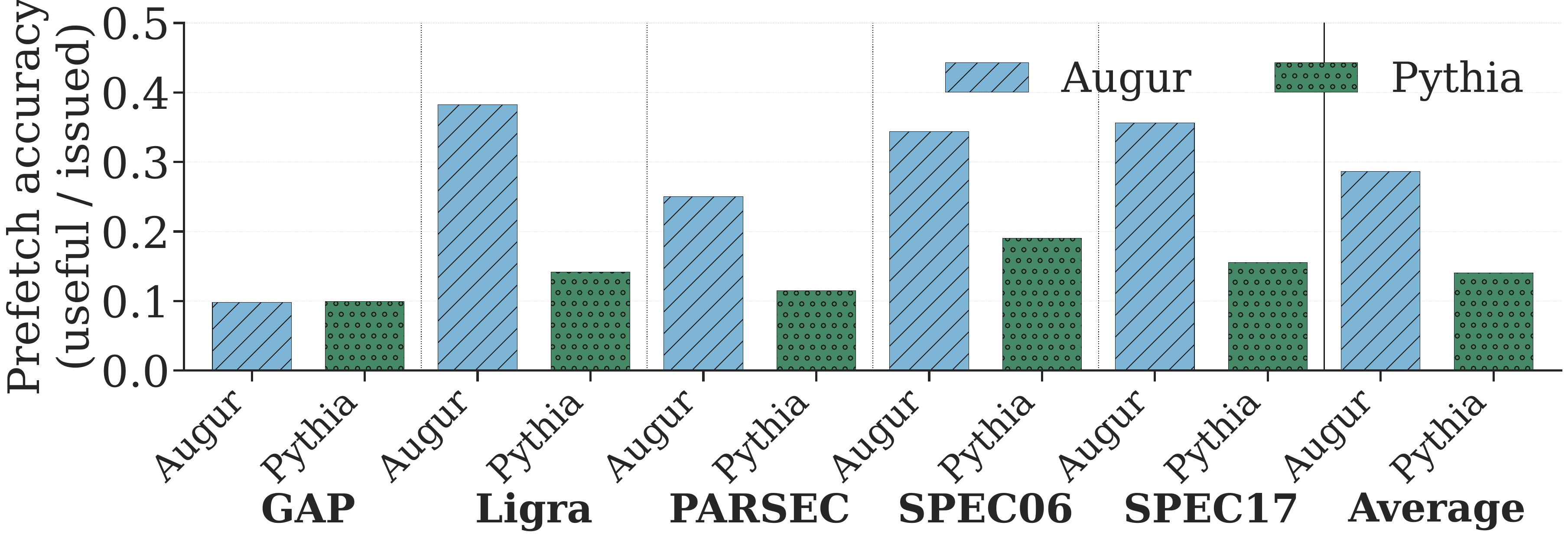}
        \caption{}
        \label{subfig:accuracy}
    \end{subfigure}
    \caption{
    Across different suites: 
    (a) Covered and overpredicted prefetches;
    (b) Prefetch accuracy for \ours~and Pythia.
    }
    \label{fig:coverage_and_acc}
    \vspace{-0.15in}
\end{figure}

\subsection{Prefetch coverage and accuracy}
Figure~\ref{subfig:coverage} shows the coverage ($=\frac{\mathrm{Useful\,prefetches}}{\mathrm{Baseline\,demand\,misses}}$) and overpredictions ($=\frac{\mathrm{Useless\,prefetches}}{\mathrm{Baseline\,demand\,misses}}$) of the L2 prefetchers across suites, and Figure~\ref{subfig:accuracy} the prefetch accuracy of \ours~and Pythia, both for a single core.
Overall, Pythia and SMS have \rev{20.8 and 18 percentage points} higher coverage than \ours, but this stems from a higher overprediction rate: \rev{37.8 and 36.4 percentage points} more overpredictions than \ours.
Averaged over all suites, \ours's prefetch accuracy is $\sim2\times$ Pythia's.


\subheading{Summary}
\ours~is more selective in issuing prefetches: it has \rev{20.8 percentage points} less coverage than Pythia, but Pythia overpredicts \rev{37.8 percentage points} more and has half \ours's prefetch accuracy.

\begin{figure}[t]
    \centering
        \centering
        \includegraphics[width=0.85\linewidth]{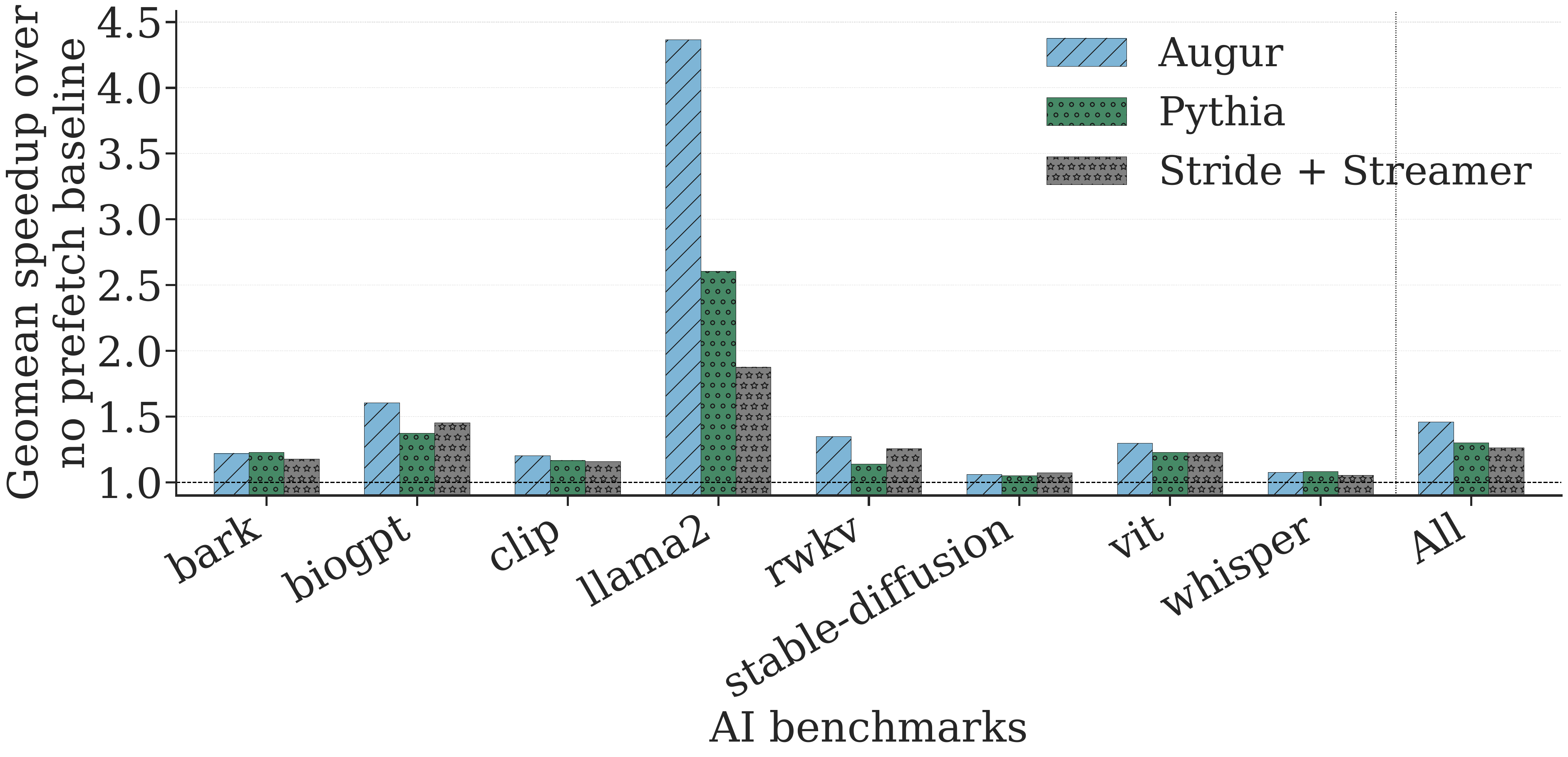}
    \caption{Speedups over a no prefetcher baseline for different prefetchers on AI workloads (single core).}
    \label{fig:additional_workloads}
    \vspace{-0.10in}
\end{figure}
\subsection{Evaluation on additional workloads}
To compare \ours~with Pythia on AI workloads, we use 32 additional traces from the Fourth Data Prefetching Competition (DPC-4)~\cite{dpc4}\rev{, from bark, biogpt, clip, llama2, rwkv, stable-diffusion, vit, and whisper}.
These traces are unseen by both Pythia (not used for tuning) and \ours~(not used during \rev{design}).
We also compare against a multi-level scheme with the Stride~\cite{stride} (L1D) and Streamer~\cite{streamer} (L2C) prefetchers found in commercial processors~\cite{Intel:ORM:2024, Viswanathan:2014:prefetcher, Sinharoy:2011:POWER7}.
Figure~\ref{fig:additional_workloads} shows single-core speedups over a no-prefetcher baseline: \ours~achieves a relative speedup of 12.2\% and 15.5\% over Pythia and Stride$+$Streamer respectively, and 67\% over Pythia on \texttt{llama2}\rev{; without \texttt{llama2}, the speedups are 6.0\% and 4.5\%}.

\subsection{Robustness to adverse system configurations}
\ours's throttling mechanisms ensure timely, selective prefetching that minimizes cache pollution, but they must be robust across system configurations.
We evaluate \ours~in adversarial configurations with limited memory bandwidth or LLC size on SPEC 2006 and SPEC 2017 (single core).

\begin{figure}[t]
    \centering
    \begin{subfigure}[b]{0.49\linewidth}
        \centering
        \includegraphics[width=\linewidth]{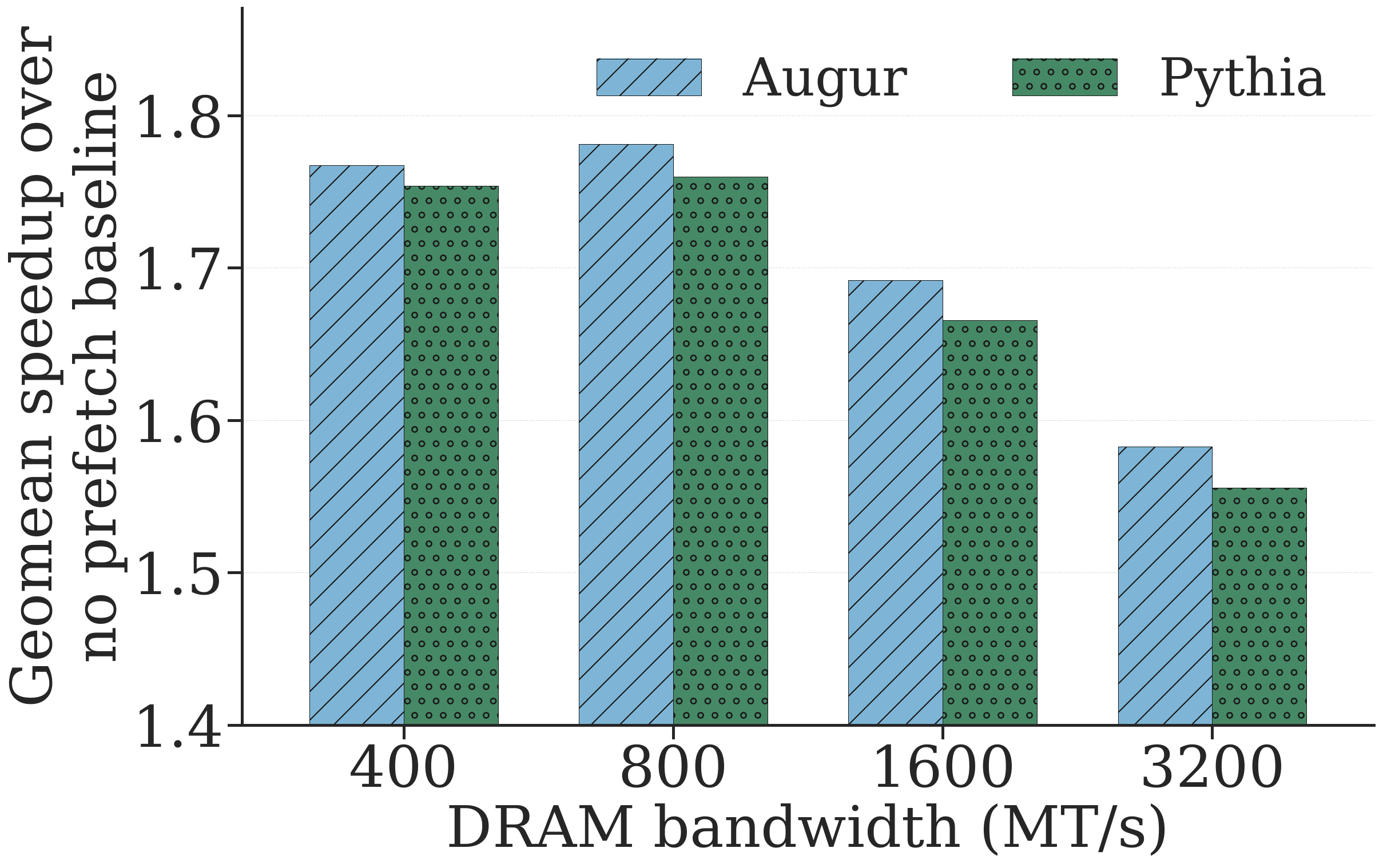}
        \caption{}
        \label{subfig:bw}
    \end{subfigure}  
    \hfill
    \begin{subfigure}[b]{0.49\linewidth}
        \centering
        \includegraphics[width=\linewidth]{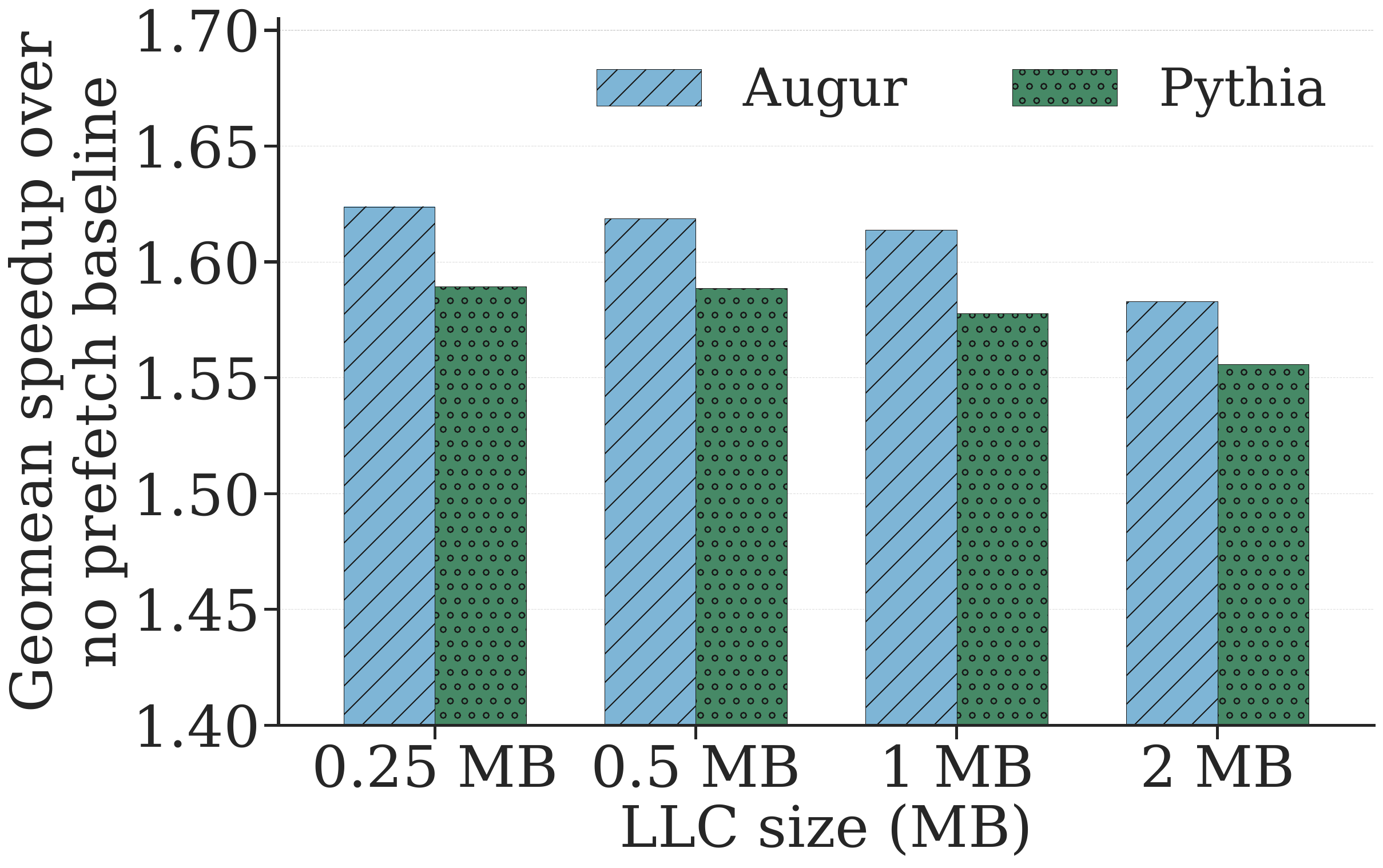}
        \caption{}
        \label{subfig:llc}
    \end{subfigure}
    \caption{Robustness of \ours~to (a) DRAM bandwidth constrained systems and (b) LLC capacity limited systems.}
    \label{fig:system_sens}
    \vspace{-0.2in}
\end{figure}

\subsubsection{Varying the memory bandwidth}
Figure~\ref{subfig:bw} shows the speedup achieved by \ours~and Pythia over different DRAM bandwidths. 
Even at low memory bandwidths of 400 MT/s, \ours~achieves a relative speedup of 0.7\% over Pythia.

\subsubsection{Varying the LLC size}
For LLC sizes of 256 KB, 512 KB, 1 MB, and 2 MB (Figure~\ref{subfig:llc}), \ours~achieves relative speedups of 2.2\%, 1.9\%, 2.3\%, and 1.7\% over Pythia.

\subheading{Summary}
We show that \ours~outperforms Pythia in both memory bandwidth and LLC capacity limited configurations without requiring any reconfiguration or tuning.

\subsection{Reservoir sensitivity studies}
We evaluate the sensitivity of \ours~to two reservoir hyperparameters, the context (history) window size and the number of tracked PCs, on SPEC 2006 and SPEC 2017 (single core).

\begin{figure}[t]
    \centering
    \begin{subfigure}[b]{0.49\linewidth}
        \centering
        \includegraphics[width=\linewidth]{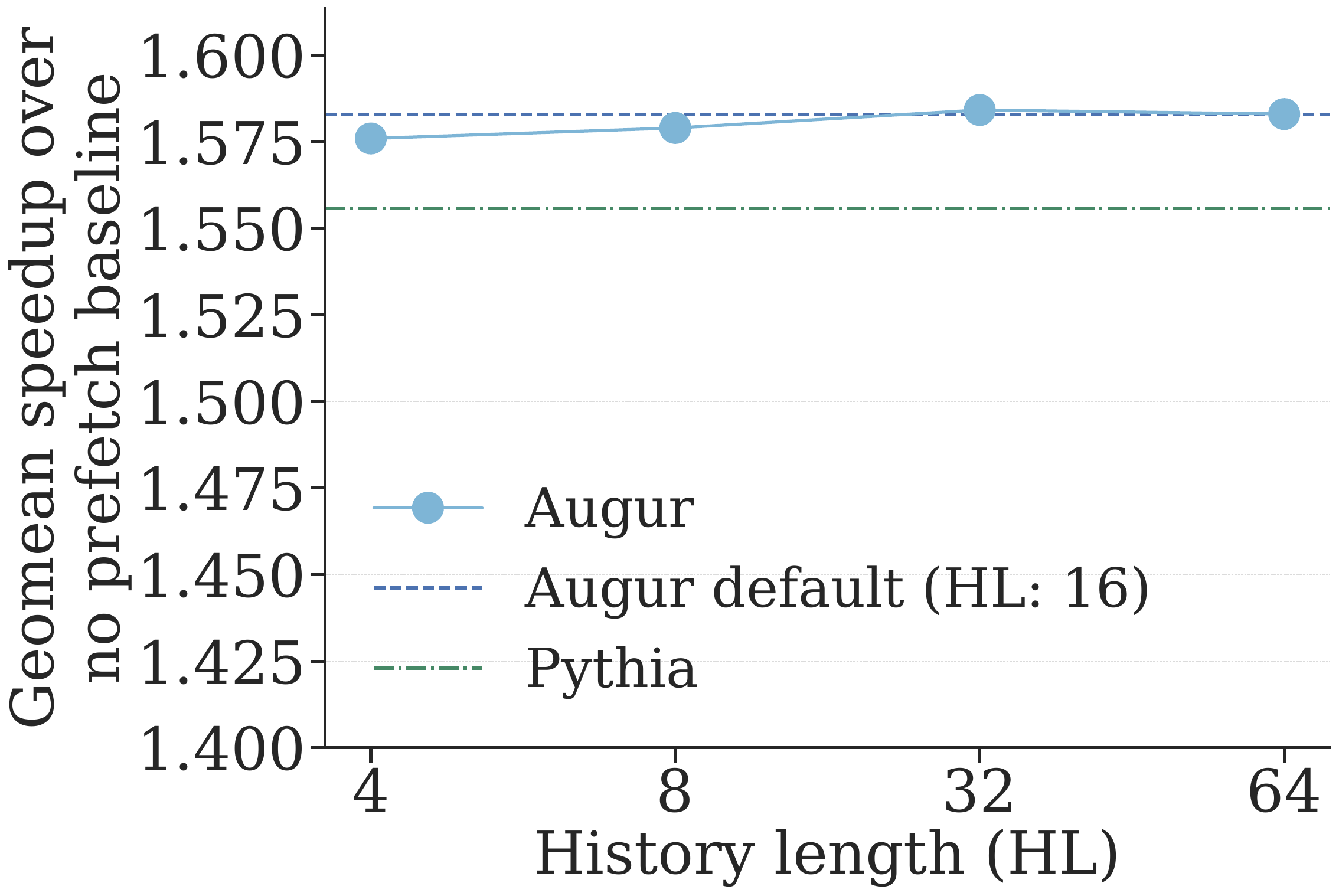}
        \caption{}
        \label{subfig:hl}
    \end{subfigure}  
    \hfill
    \begin{subfigure}[b]{0.49\linewidth}
        \centering
        \includegraphics[width=\linewidth]{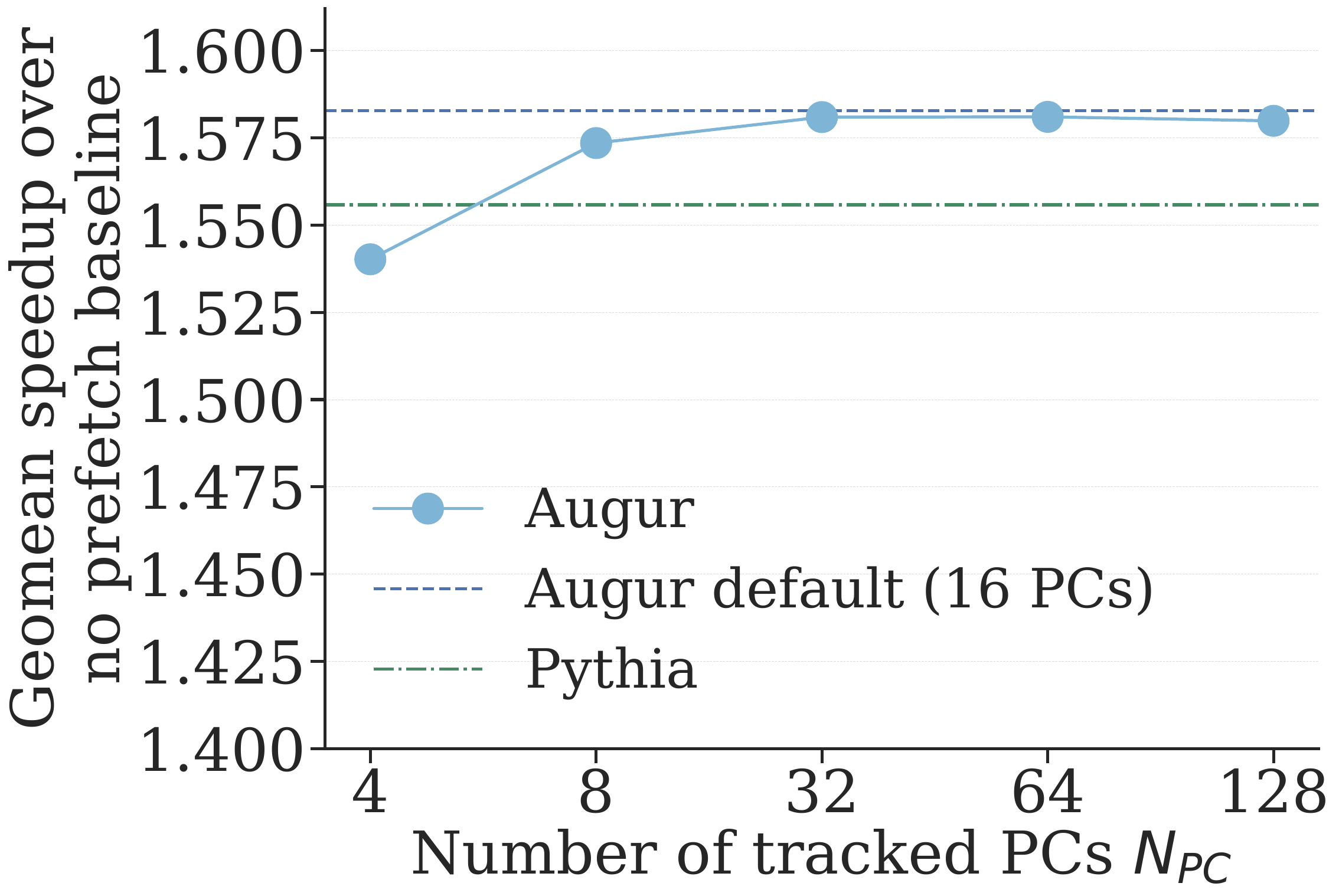}
        \caption{}
        \label{subfig:pcs}
    \end{subfigure}
    \caption{
    Sensitivity of \ours~to reservoir configurations with different (a) history lengths (HL) and (b) tracked PCs ($N_{PC}$) \rev{(SPEC; y-axis starts at 1.4)}.
    }
    \label{fig:res_sens}
    \vspace{-0.2in}
\end{figure}

\subsubsection{Varying the reservoir history length}
The history length (HL) sets the context of past (global) demand accesses seen by \ours; the base \ours~uses $HL = 16$.
Figure~\ref{subfig:hl} varies HL from small ($HL = 4$) to large ($HL = 64$): a larger HL captures inter-PC dependencies, but the gain over 4 entries is marginal.

\subsubsection{Varying the number of tracked PCs}
Figure~\ref{subfig:pcs} varies the number of tracked PCs ($N_\mathrm{PC}$).
With $N_\mathrm{PC}=4$, \ours~cannot prefetch for enough PCs to compete with Pythia, but with $N_\mathrm{PC}=8$ it achieves a relative speedup of 1.1\% over Pythia.

\subheading{Summary}
We show that even when the reservoir context window history length is reduced to 4, \ours~achieves a relative speedup of 1.3\% over Pythia \rev{on SPEC (6.7\% on GAP+Ligra, Section~\ref{sec:learning})}. 

\subsection{Ablation studies}
We now show the results of two ablation studies: (1) \ours~without the Markov predictor (\ours-R), and (2) A cost-equivalent baseline with a larger L2C (Baseline++). 

\begin{figure}[t]
    \centering
    \begin{subfigure}[b]{0.85\linewidth}
        \centering
        \includegraphics[width=\linewidth]{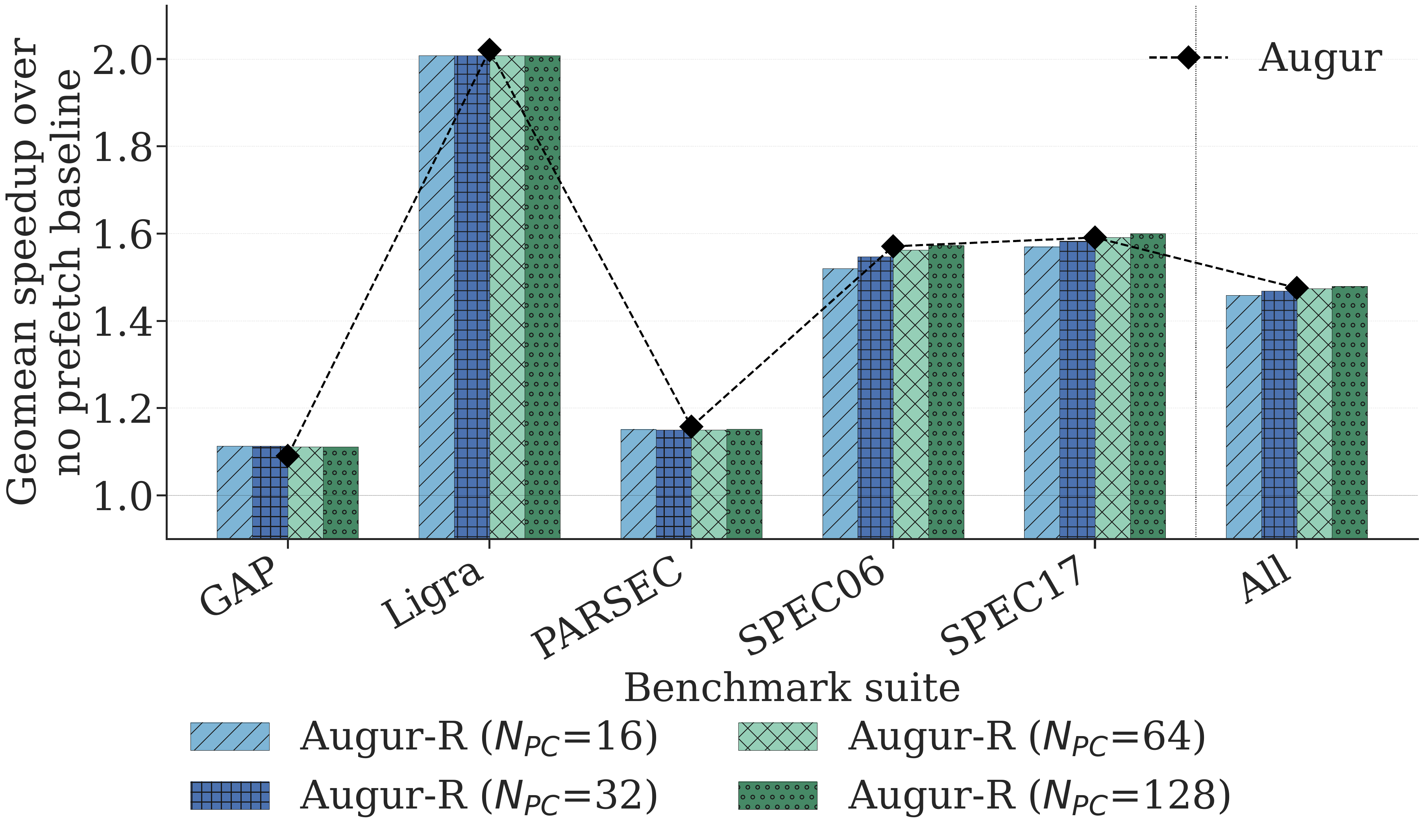}
        \caption{}
        \label{subfig:nomarkov}
    \end{subfigure}  
    \\
    \begin{subfigure}[b]{\linewidth}
        \centering
        \includegraphics[width=0.85\linewidth]{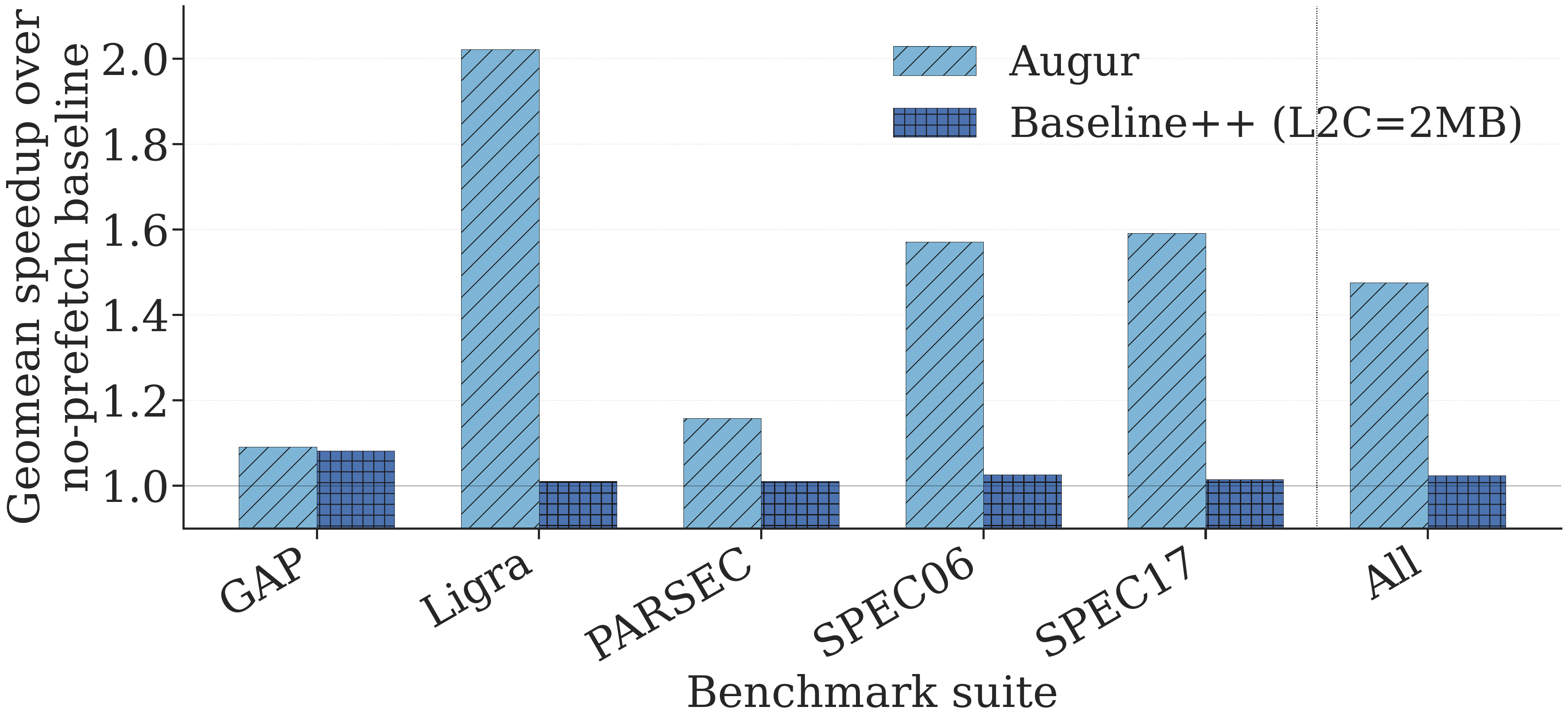}
        \caption{}
        \label{subfig:cost-equiv}
    \end{subfigure}
    \caption{
    Geo-mean speedups for (a) \ours-R (no Markov predictor); (b) \ours~over Baseline++ \rev{(y-axes do not start at 0)}.
    }
    \label{fig:ablation}
    \vspace{-0.2in}
\end{figure}

\subsubsection{\ours~without the Markov predictor}
Figure~\ref{subfig:nomarkov} shows single-core performance of \ours-R for different $N_{PC}$ (HL=16).
Removing the Markov predictor has no effect on PARSEC, or on GAP and Ligra, whose PCs rarely have trivial patterns (Section~\ref{sec:sources}).
The SPEC suites, however, require higher $N_{PC}$ ($>32$) for \ours-R to match \ours, as SPEC benchmarks have many PCs with constant (invariant) or simple alternating deltas.
Since the NGRC must track at least 64 PCs to match \ours, the lightweight Markov predictor significantly reduces the NGRC's cost; it predicts no patterns the NGRC cannot, but simply filters trivial delta patterns.

\subsubsection{Cost-equivalence study}
Baseline++ spends \ours's area on a larger L2C instead of a prefetcher; we optimistically assume this doubles the L2C (1MB to 2MB).
Figure~\ref{subfig:cost-equiv} compares \ours~and Baseline++ against the no-prefetch 1MB-L2C baseline (single core).
Baseline++ slightly improves over this baseline, but \ours~outperforms it by 44\% across all suites.

\subsection{\rev{Where does \ours's benefit come from?}}\label{sec:sources}

\begin{figure}[t]
  \centering
  \includegraphics[width=\columnwidth]{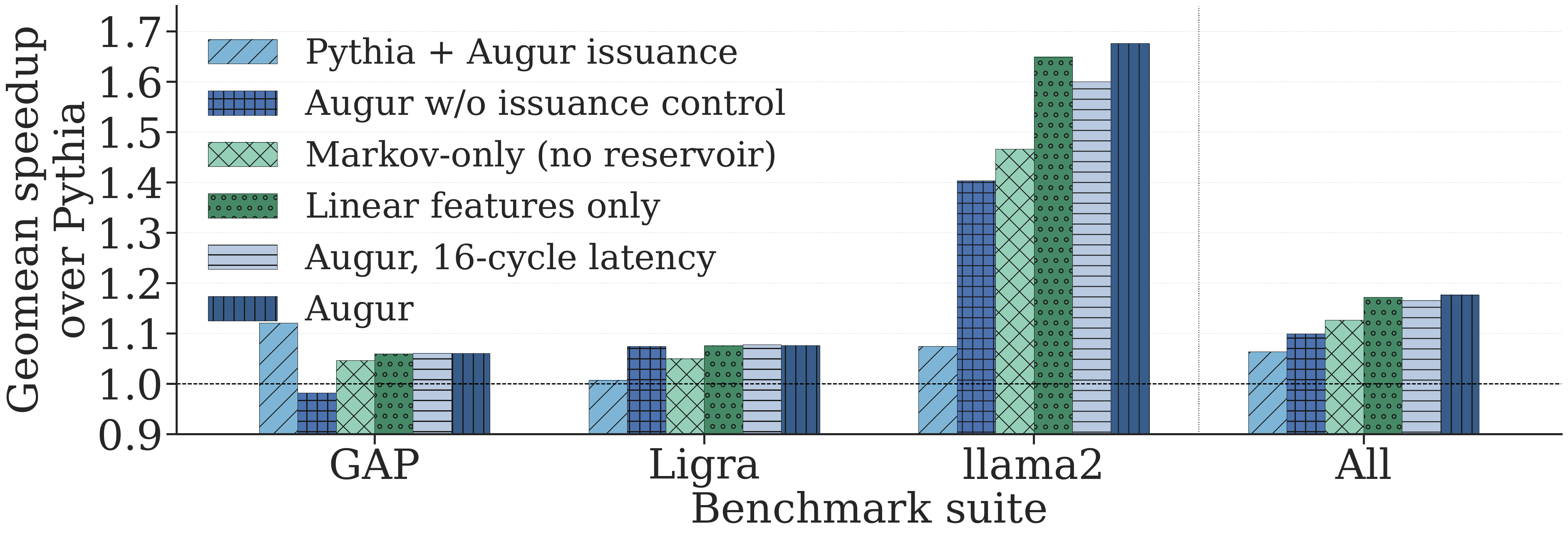}
  \caption{\rev{Component isolation: geomean speedup over Pythia (``All'' = the 42 GAP, Ligra, and llama2
  traces).}}
  \label{fig:components}
\end{figure}

\rev{To understand where \ours's speedup comes from, we isolate each of its components on the GAP, Ligra, and \texttt{llama2} traces (Figure~\ref{fig:components}). All results come from one design, which reproduces every single-core IPC in Figure~\ref{fig:speedups} exactly.}

\rev{\textbf{Prediction vs.\ issuance.} \ours's speedup comes from two sources: a better predictor, and more selective issuance.
When Pythia uses \ours's issuance controls, it improves by 6.4\%, even though its RL reward already makes it selective, and \ours~still outperforms it by 10.5\%.
So about 60\% of \ours's advantage comes from prediction, and the rest from selectivity.
The split depends on the suite: on GAP, Pythia with our issuance controls outperforms \ours, while on Ligra and \texttt{llama2} almost all of the gain comes from prediction.
Without its issuance controls, \ours~is only 9.9\% faster than Pythia.
Replacing the reservoir with the Markov predictor reduces performance by 4.3\% (12.5\% on \texttt{llama2}).}

\rev{\textbf{Nonlinear features.} Using AND instead of multiplication to generate the nonlinear features has no measurable cost (0.2\%).
A linear-only readout also comes close: over all 169 traces, it is only 0.3\% slower than \ours. NGRC's time-delay embedding does most of the work.
The nonlinear features help most on traces where the next delta depends on the combined context (e.g., streamcluster, rwkv, gcc, and \texttt{llama2}), by up to 4.8\%.
Since a linear-only \ours~needs only $\sim$14~KB, the nonlinear expansion can be dropped when storage is tight.}

\rev{\textbf{Prefetch sources.} Table~\ref{tab:sources} breaks down the prefetches issued by each source.
Overall, the reservoir issues 44\% of \ours's prefetches but supplies 59\% of its useful ones.
On GAP, most prefetches come from the next-line fallback, and disabling it improves GAP by 7.3\% (our GAP results are conservative). We keep it since it supplies 14--20\% of useful prefetches on PARSEC and SPEC 2006, where PCs churn.
On \texttt{llama2}, 98\% of per-PC deltas are $+1$, and the reservoir supplies almost all useful prefetches at 91\% accuracy.
Pythia's accuracy on \texttt{llama2} is only 19\%, mostly because 76\% of its prefetches are for lines that are already being fetched. Our issuance controls improve Pythia by only 7.4\% here, so the gap is in prediction.
Note that accuracies in this section are pooled over the traces evaluated here, while Figure~\ref{subfig:accuracy} averages per suite.}

\begin{table}[t]
  \caption{\rev{Issued/useful share (accuracy) by source; late: \ours{}/Pythia (\%)}}
  \label{tab:sources}
  \centering
  \footnotesize
  \setlength{\tabcolsep}{3.5pt}
  \rev{\begin{tabular}{lccccc}
    \toprule
    Suite & Reservoir & Markov & NL fallback & NL warm-up & Late \\
    \midrule
    GAP    & 19/29 (77) & 7/3 (22)   & 74/68 (46) & 0/0   & 3/8   \\
    Ligra  & 58/83 (79) & 7/2 (17)   & 35/15 (23) & 0/0   & 6/14  \\
    PARSEC & 49/63 (49) & 19/22 (45) & 27/14 (19) & 5/1   & 7/25  \\
    SPEC06 & 35/47 (70) & 30/31 (53) & 30/20 (35) & 6/2   & 10/13 \\
    SPEC17 & 41/61 (69) & 29/32 (52) & 12/4 (17)  & 19/3  & 8/18  \\
    AI     & 69/79 (65) & 26/20 (42) & 4/2 (24)   & 1/0   & 49/59 \\
    \midrule
    All    & 44/59 (69) & 26/26 (50) & 20/13 (32) & 10/2  & 18/23 \\
    \bottomrule
  \end{tabular}}
\end{table}

\subsection{\rev{Online learning, configuration, and timeliness}}\label{sec:learning}

\begin{figure}[t]
  \centering
  \subfloat[\rev{Convergence}]{\includegraphics[width=0.49\columnwidth]{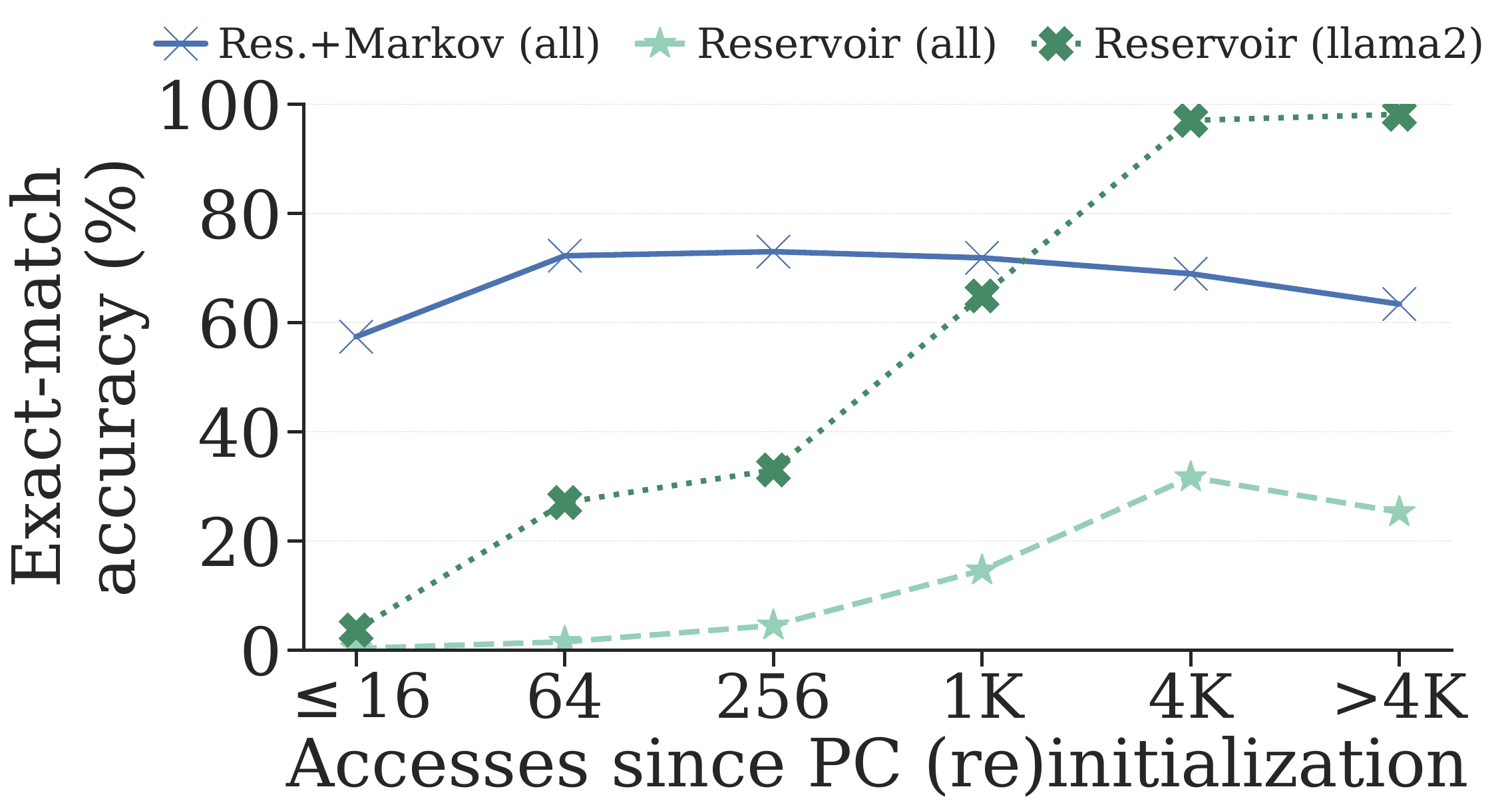}}\hfill
  \subfloat[\rev{Per-trace speedup}]{\includegraphics[width=0.49\columnwidth]{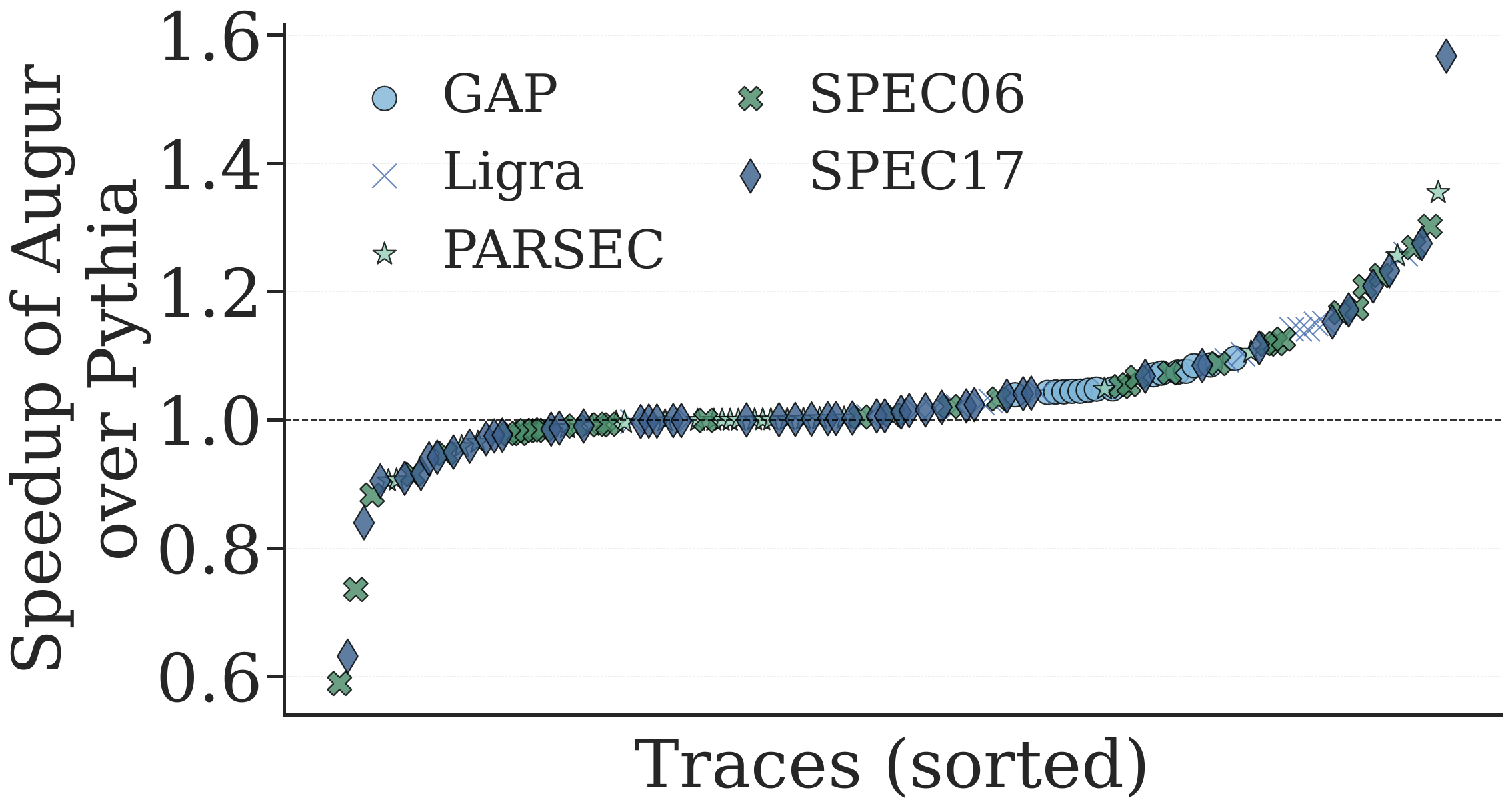}}
  \caption{\rev{(a) Exact-match accuracy vs.\ accesses since a PC was (re)initialized; (b) per-trace
  speedup of \ours{} over Pythia (137 single-core traces, sorted).}}
  \label{fig:learning}
\end{figure}

\rev{\textbf{Online learning.} Figure~\ref{fig:learning}(a) shows how quickly \ours~learns after a PC is (re)initialized.
The reservoir needs a few thousand accesses per PC to converge, but the Markov predictor covers this period, and 57\% of \ours's predictions are already correct within a PC's first 16 accesses.
PC churn is mostly limited to SPEC and PARSEC, where 9--25\% of accesses come from recently (re)initialized PCs; on GAP and Ligra, tracking 8 or 32 PCs instead of 16 makes no difference.
Median load PCs per trace range from 26 (Ligra) to 180 (SPEC 2006), and median L2 MPKI from 0.6 (PARSEC) to 48 (GAP).}

\rev{\textbf{Sensitivity and storage.} We also vary the learning rate and its schedule, the Markov margin, the confidence threshold, the history length, and the bias on GAP and Ligra; none of them changes performance by more than 0.5\%.
Only a very small base degree (2) hurts, by 2.8\%.
These values were set during development on SPEC and GAP and were not tuned per benchmark; Ligra, PARSEC, and the AI traces were only evaluated with the final design.
With a history length of 8 (17.2~KB) or 4 (13.0~KB), \ours~still outperforms Pythia by 6.8\% and 6.7\% on GAP and Ligra, and by 1.5\% and 1.3\% on SPEC.
In contrast, growing Pythia to 35.3 or 49.5~KB changes its performance by at most 0.6\%.}

\rev{\textbf{Filtering and timeliness.} The prefetch filter and the usefulness throttle both matter: removing either one reduces performance by $\sim$3.7\% on GAP and Ligra, while the PQ/bandwidth caps and adaptive lookahead have little effect.
Throttling does suppress some useful prefetches (36\% of throttled predictions are later missed), but only 0.5\% of the prefetches dropped by the filter are later missed.
18\% of \ours's useful prefetches are late (Pythia: 23\%).}

\rev{\textbf{Prediction latency.} Like DPC-4, our simulator does not model the latency of any prefetcher.
We add a pipelined datapath where a prediction takes 4--32 cycles and its weight update lands only when it completes, so back-to-back accesses from one PC use stale weights.
With a 16-cycle latency, \ours~still outperforms Pythia by 3.1\% on the 137 single-core traces (2.9\% without latency; the delay happens to help a few SPEC 2006 traces).
Only \texttt{llama2}, whose accesses are very close together, is sensitive to latency: its speedup over Pythia drops from 67.6\% to 60.0\% at 16 cycles.}

\rev{\textbf{Multi-core metrics, simulation length, and per-trace results.} \ours's weighted speedup over Pythia is 5.5\% and 6.1\% on four and eight cores, slightly higher than the throughput speedups in Figure~\ref{fig:speedups}; the harmonic speedup is on par with Pythia, so the gain is not shared evenly across cores.
Longer simulations do not shrink \ours's advantage: on GAP, a 400M-instruction run or a 200M warm-up raises it from 6.0\% to 6.2\%, and on four-core Ligra, a 50M+200M run per core raises it from 7.6\% to 9.2\%.
Of the 137 single-core traces, \ours~wins on 72, ties on 31, and loses on 34, mostly mcf (Figure~\ref{fig:learning}(b)); without its five best traces, it is still 1.9\% faster.}

\subsection{Hardware overheads}

We implemented \ours~in Verilog and synthesized the design using Cadence Genus~\cite{genus} and a commercially available 12\,nm technology library. A single instance of \ours~for a private L2C per core occupies 0.32~mm$^2$ and consumes 32.7~mW \rev{(synthesized with all 560 weights per PC, i.e., before merging the square-term weights)}. Our implementation does not rely on an SRAM compiler; instead, all storage structures are realized using synthesized logic. As a result, the reported area should be interpreted as a conservative upper bound. If SRAM compiler macros were used, we expect the overall area to be significantly smaller due to the higher density of compiled memory cells. 
For context, this 0.32~mm$^2$ footprint is small compared to the L2 caches of contemporary commercial processors, whose capacities (across varying L2 topologies) typically range from the order of \textit{1 MB} (e.g., Intel, AMD) to 32 MB (e.g., IBM), although exact L2 area depends on process technology and design.

%% file: inputs/9RelatedWorks.tex
\section{Related Work}
To the best of our knowledge, \ours~is the first delta prefetcher based on next-generation reservoir computing. 
Since \ours~is a delta prefetcher, it is orthogonal to works on temporal prefetchers. 

\subheading{Heuristic delta prefetchers}
These predict the delta between consecutive memory accesses, globally or per PC, and range from simple stride detectors to prefetchers that compress stride histories into signatures to identify complex patterns~\cite{jouppi_prefetch, baer2, stride, streamer, stride_vector, ampm, dspatch, sms_mod, sms, footprint, fdp, spp, ipcp, vldp, bingo, mlop, dol, bop, navarro2022berti,dmp,navarro2025complexity,mutlu2005address,sandbox,domino}.

\subheading{Machine learning in microarchitecture research}
ML has been widely applied to tasks such as branch prediction~\cite{tarsa,branchnet,perceptron,garza2019bit,BranchNetV2_2020,Zouzias2021BranchPA,jimenez2016multiperspective,tarjan2005merging,jimenez2003fast,jimenez2002neural}, memory scheduling~\cite{rlmc,morse}, cache management~\cite{glider, imitation,rl_cache,teran2016perceptron,balasubramanian2021accelerating}, address translation~\cite{margaritovTranslationNIPS18}, off-chip prediction~\cite{hermes,jamet2024tlp,athena_extended}, prefetch filtering~\cite{vavouliotis2025cross,ppf,yang2025reinforcement}, and hardware data prefetching~\cite{peled2018neural,peled_rl,hashemi2018learning,shineural,shi2019learning,Shi2021,zeng2017long,duong2024new,pythia,zhang2022resemble,pathfinder,Srivastava2020,Zhang2020}.
\rev{Closest in mechanism, PPF~\cite{ppf} trains a perceptron online to filter SPP's prefetches, and hashed and multiperspective perceptrons~\cite{tarjan2005merging,jimenez2016multiperspective} classify using weights indexed by hashed features; \ours~instead uses online regression over a time-delayed window of hashed features as the predictor itself. Micro-Armed Bandit~\cite{mab} applies lightweight online learning to prefetcher control rather than prediction, and Gaze~\cite{chen2025gaze} is a spatial prefetcher, complementary to delta prediction.}

\subheading{ML prefetchers}
Voyager~\cite{Shi2021} and Twilight~\cite{duong2024new} are temporal prefetchers that use attention-based embeddings in their formulation and are orthogonal to \ours.
LSTM- and RNN-based approaches~\cite{hashemi2018learning,zeng2017long,Zhang2020} are significantly more expensive to implement on hardware than \ours, and must be trained offline which inhibits their generality.
This is applicable to neural network-based approaches~\cite{shineural, peled2018neural} as well. 
Only Pathfinder~\cite{pathfinder}, a spiking neural network based prefetcher, and Pythia~\cite{pythia}, a reinforcement learning based prefetcher, have been shown to be practical for hardware implementations.
However, Pathfinder requires complex training protocols, and Pythia required hundreds of offline simulations to train its feature and actions sets. 
In contrast, \ours~has a simple, online, training loop, and generalizes across benchmark suites \rev{without per-benchmark tuning} to outperform Pythia.

%% file: inputs/10Conclusions.tex
\section{Conclusions}

In this paper, we present \ours, a practical, fully online, ML-based delta prefetcher for L2 caches. 
\ours~applies key microarchitectural optimizations to NGRC so that it can be trained entirely online, \rev{including bitwise AND feature generation and integer-only training and inference}, \rev{and pairs it with a selective issuance policy that also improves other prefetchers}. \ours~outperforms the state-of-the-art prefetcher Pythia by \rev{2.9\%, 4.5\%, and 5.3\% on 1-, 4-, and 8-core systems} with very small hardware overheads (0.32~mm$^2$ area, 32.7~mW power, \rev{33~KB} of memory), while training fully online \rev{without per-benchmark tuning}. Our results open the door to a new paradigm of AI-infused microarchitecture that is practical to implement in hardware, with opportunities to extend this work to other speculative processor components, including branch and hit/miss prediction.